\documentclass[aps,floats,pre,showpacs,twocolumn
]{revtex4}

\newcommand{\hL}{{\hat {\cal L}}'_{12}}
\newcommand{\V}{{\bf  V}}
\newcommand{\C}{{\bf  C}}

\newcommand{\G}{{\bf \Gamma}}
\newcommand{\E}{{\hat {\cal E}}}
\newcommand{\U}{{\hat {\frak U}}}
\newcommand{\Li}{{\hat {\frak L}}}
\newcommand{\GG}{{\hat {\frak g}}}
\newcommand{\R}{{\hat {\frak R}}}
\newcommand{\LI}{{\hat {\cal L}}}

\usepackage{amsfonts,amsmath} \usepackage{bm} \usepackage{dcolumn}
\usepackage{epsfig} \usepackage{latexsym}

\begin{document}

\title{Manifestly covariant classical correlation dynamics I. General theory}

\author{Chushun Tian}
\affiliation{Institut f{\"u}r Theoretische Physik, Z{\"u}lpicher
Str. 77, K{\"o}ln, D-50937, Germany}

\date{\today}
\begin{abstract}
 {\rm In this series of papers we substantially
 extend investigations of Israel and Kandrup on nonequilibrium statistical
 mechanics in the framework of special relativity. This is the first
 one devoted to the general mathematical structure.
 Basing on the action-at-a-distance formalism we obtain a
 single-time Liouville equation. This equation describes the manifestly covariant evolution of
 the distribution function of full classical many-body systems. For such global evolution the
 Bogoliubov functional assumption is justified. In particular, using
 the Balescu-Wallenborn projection operator approach we find that
 the distribution function of full many-body systems is
 completely determined by the reduced one-body distribution function. A manifestly covariant closed nonlinear
 equation satisfied by the reduced one-body distribution function is rigorously derived. We also discuss
 extensively the generalization to the general relativity especially an application to self-gravitating systems.}
\end{abstract}

\maketitle

\section{Introduction and overview}
\label{introduction}

\subsection{Some motivations}
\label{motivation}

The fundamental problem of physics of relativistic classical
many-body systems is one century old (for a review of early
investigations see, e.g., Ref.~\cite{Hakim67}) and, still, remains
one of the most important subjects in studies of relativity
\cite{Schieve05,Hanggi07,Hanggi05,Kirpichev07}. At the microscopic
level the basis of deterministic classical dynamics of relativistic
many-body systems has been widely explored
\cite{Fokker29,Wheeler49,Dirac49,Martinez06}. At the macroscopic
level relativistic hydrodynamical phenomena have been well
documented \cite{Cercigani02,vanLeeuwen80}. In particular, in recent
years relativistic effects on macroscopic equilibrium and
non-equilibrium phenomena have attracted considerable theoretical
interests, and a variety of related subjects are under intensified
investigations ranging from the internal evolution (or aging) of
relativistic systems \cite{Yaacov06,Yaacov95}, relativistic
generalization of the Maxwell-Boltzmann distribution
\cite{Hanggi07,Debbasch08} to relativistic Brownian motion
\cite{Hanggi05}. A natural question, thus, is {\it how various
macroscopic relativistic phenomena emerge from deterministic
relativistic classical many-body dynamics}. Long time ago it was
realized by the Brussels-Austin school that, as nonrelativistic
many-body systems, the macroscopic evolution generally roots in the
creation or annihilation of correlations between microscopic
particles as they move in spacetime \cite{Prigogine65,Balescu64}.
Thus, to investigate these issues one has to invent a theory of {\it
relativistic classical correlation dynamics} aiming at
understandings of the dynamics of distribution function of full
many-body systems. Unfortunately, this is intellectually challenging
and remains largely undeveloped.

From practical viewpoints a theory of relativistic classical
correlation dynamics may be proven to be powerful in studies of
various relativistic transport processes in plasma physics
\cite{Lefebvre08} and astrophysics \cite{Israel84,Kandrup84}. There
substantial progresses have been made in past years by making use of
various relativistic kinetic equations, which provide an adequate
microscopic description for macroscopic relativistic hydrodynamics.
In a pioneering paper by Heinz \cite{Heinz83} it was realized that
the relativistic classical kinetic equation may also become a
powerful tool in studies of quark-gluon plasmas, which has nowadays
spanned one of the most important branches in high-energy physics
\cite{Litim02}. To derive relativistic (classical) kinetic equations
there are a number of theoretical proposals. Among of them are two
representative approaches: The first approach is largely
phenomenological, where with the help of {\it Stosszahlansatz}
kinetic equations are formulated as manifestly covariant versions of
their Newtonian analogues so as to suit relativity principles
\cite{Cercigani02,vanLeeuwen80}. The second approach
\cite{Naumov81,Schieve89,Lu94,Tian01} starts from some relativistic
BBGKY hierarchy and the cluster expansion is further applied. Then,
in the crucial step one invokes the Bogoliubov functional assumption
\cite{Bogoliubov} to truncate the infinite hierarchy. As such, one
obtains some relativistic kinetic equation. Yet, there does exist
the notable exception in plasma physics, namely Klimontovich's
technique. With the help of this technique Klimontovich managed to
justify the relativistic Landau equation \cite{Belyaev57} at the
full microscopic level \cite{Klimontovich60,Klimontovich67}
. Since then, it has become common to test the validity of various
theories by justifying this kinetic equation
\cite{Hakim67,Prigogine65,Naumov81,Lu94,Israel84,Kandrup84}.

It is fair to say that for some important applications such as
ultrarelativistic electromagnetic plasmas the particle interaction
must be described by the quantum field theory. Therefore, a complete
treatise of relativistic statistical effects is required to be built
on quantum field theories. Then, a general prescription may be to
generalize--in a manifestly covariant manner--the Green's function
theory of nonequilibrium quantum statistical mechanics, pioneered by
Schwinger, Kadanoff and Baym, Keldysh and Korenman
\cite{Schwinger,Kadanoff} so as to suit principles of quantum field
theories. Such a task was undertaken for electron-positron plasmas
some time ago \cite{DuBois72}. The general formalism has been
substantially renovated by making use of the functional-integral
approach, emerging as relativistic nonequilibrium quantum field
theory \cite{Chou85,Hu88}. These approaches, in combination with the
Wigner function technique, allow one to obtain a manifestly
covariant quantum kinetic equation at the level of the weak coupling
(Born) approximation which, importantly, recovers the relativistic
Landau equation \cite{Belyaev57,Klimontovich60} at the classical
limit.

Nevertheless there are considerable practical and theoretical
reasons which justify to intensify studies of the relativistic
classical correlation dynamics. The most prominent one comes the
very recent debate on the special relativistic generalization of the
Maxwell-Boltzmann distribution \cite{Hanggi07,Debbasch08}, triggered
by the surprising finding of Horwitz and coworkers \cite{Schieve89}.
In Ref.~\cite{Schieve89} it was noticed that considerable conceptual
and technical differences between the relativistic and Newtonian
deterministic classical dynamics might completely disable simple
generalizations obtained by the phenomenological approach. These
authors proposed a manifestly covariant kinetic equation possessing
a remarkable mathematical structure different from the one obtained
either by other microscopic approaches
\cite{Prigogine65,Lu94,Klimontovich60,Israel84,Kandrup84,DuBois72,Chou85,Hu88}
or by phenomenological arguments \cite{Cercigani02,vanLeeuwen80}.
The one century old belief--that the J{\"u}ttner equilibrium
\cite{Hanggi07,Debbasch08} is established in relativistic dilute
systems--is thereby challenged. At this stage it is completely
unclear whether such a peculiar property is specific or universal to
relativistic classical many-body systems. Thus, it is fundamentally
important to derive as rigorously as possible the relativistic
kinetic equation from generic classical many-body dynamics. Then,
one may hope to prove or disprove kinetic equations proposed basing
on phenomenological arguments and, furthermore, to obtain insight
into the numerical experiment on the relativistic equilibrium of
{\it classical} many-body systems \cite{Hanggi07}. Experiences in
the Newtonian statistical mechanics suggest that the technique
developed by the Brussels-Austin school \cite{Prigogine63,Balescu}
may be perfectly suitable to fulfill such a task as it manages to
derive a general kinetic equation \cite{kinetic} without resorting
to either {\it Stosszahlansatz} or the truncation approximation. To
employ the latter is well-known to be inevitable and, in fact, a
crucial step in various microscopic approaches.

There are several important adjacent problems that may be explored
within the scope of the relativistic classical correlation dynamics.
One is the relativistic Brownian motion, which has received much
attention in recent years (for example, see \cite{Hanggi05}).
Another is to formulate the relativistic many-body equilibrium. For
this Hakim conjectured long time ago that there exists a Lorentz
invariant equilibrium hierarchy which is invariant under the
spacetime translation and is merely determined by the J{\"u}ttner
distribution \cite{Hakim67}. However, to our best knowledge there
have been no progresses reported so far.

Finally, new developments of the relativistic classical correlation
dynamics are urged by the stellar dynamics. There, it is generally
believed that a satisfactory framework is provided by the classical
(nonequilibrium) statistical mechanics in the general relativity
\cite{Israel84}. In particular, the kinetic equation describing the
Brownian motion of self-gravitating systems suffers from a serious
difficulty of the infrared divergence
\cite{Chandrasekhar,Israel84,Kandrup86}. Such a divergence has been
explored at the level of the Newtonian physics and been known to
deeply root in the nonMarkovian and collective effect
\cite{Prigogine66,Kukharenko94}, the latter of which, interestingly,
finds the collective dielectric effect in classical plasmas
\cite{Balescu60} as the analogue. In the Newtonian physics the
correlation dynamics encompasses a standard route towards the
complete treatise of these effects. It is, therefore, a natural hope
that the (general) relativistic classical correlation dynamics may
help to understand this issue at deeper level.

\subsection{Nonequilibrium statistical mechanics:\\
manifestly and nonmanifestly covariant formalism}
\label{covariant}

In this series of papers we focus on the (special relativistic)
classical statistical mechanics. In contrast to the relativity in
the Newtonian physics the emergence of macroscopic irreversibility
from deterministic classical dynamics nowadays is a well established
subject. Indeed, for simple chaotic systems especially deterministic
diffusive systems rigorous mathematical investigations, as well as
various analytical theories \cite{Gaspard96} have shown that an
initial phase space distribution (function) is decomposed into
independent components each of which relaxes in an irreversible
manner (Ruelle-Pollicott resonances), and after transient processes
the evolution is dominated by diffusion at the macroscopic scale.
For more complicated many-body systems (as we switch to in this
paper) a milestone was put by Bogoliubov \cite{Bogoliubov} who
realized that, there, the evolution is a two-step process: In the
first step (after transient decays) correlation functions relax into
functionals of one-body distribution functions. In the second step
the one-body distribution function evolves following a closed
nonlinear kinetic equation, and is fully responsible for macroscopic
hydrodynamics. Later on the concept of the Bogoliubov functional
assumption was justified and substantially extended, especially in
the profound work by the Brussels-Austin school using delicate
diagrammatical \cite{Prigogine63} and projection operator
\cite{Balescu,Prigogine96} approaches. In particular, the
decomposition of many-body phase space distribution functions may be
carried out in the way such that in the absence of interactions each
component corresponds to some degree of correlations between
particles. These correlations are preserved (mixed) in the absence
(presence) of interactions justifying the terminology of
``(classical) correlation dynamics''. As such, irreversible
processes of approaching equilibrium, as well as the final
equilibrium states are found to be completely determined by the
so-called vacuum of correlations namely the component containing no
correlations in the absence of interactions. This way the kinetic
equation is no longer an approximation, rather, an exact theory to
which the Liouville equation of full many-body systems converges in
long times.

Then, a natural hope is to extend these concepts so as to suit
relativity principles. At the early stages the Brussels-Austin
school \cite{Prigogine65,Balescu67} undertook systematic attempts
(see Ref.~\cite{Yaacov95} for a review) to generalize the Newtonian
classical correlation dynamics to special relativity, which is built
on the canonical formulation of relativistic dynamics of classical
many-body systems \cite{Dirac49} with the Hamiltonian, if necessary,
including field degrees of freedom. Unfortunately, as pointed out by
many authors especially in the notable critical analysis by Hakim
\cite{Hakim67}, Israel and Kandrup \cite{Israel84}, and Kandrup
\cite{Kandrup84}, to proceed along this theoretical line one may
have to overcome a number of conceptual and technical difficulties.
The most serious problems are regarding {\it covariance of the
theory} and particularly {\it covariance of the evolution of
many-body systems}. Both deeply root in the absence of absolute time
in many-body systems. For the former problem it should be stressed
that by relativity principles it is perfectly legitimate to build
any theories in either manifestly or non-manifestly covariant
manner. It is exactly the latter that the Brussels-Austin school
follows. Indeed, in the development of old relativistic classical
correlation dynamics a preferred time coordinate is chosen and then
a Liouville equation is formulated building on the Hamiltonian
formalism. Furthermore, in an insightful work Balescu and Kotera
realized that in this framework the Lorentz invariance must be
understood in terms of the Lorentz group action on the distribution
functions of many-particle phase space (if necessary, enlarged to
accommodate field degrees of freedom) \cite{Balescu67,Yaacov95}.
More precisely, the Lorentz group has $10$ generators and in the
group action representation the Liouvillian generates the time
translation. Then, it is crucial to justify that kinetic equations
obtained in subsequent calculations (which require further delicate
approximations) also follow this kind of Lorentz invariance. Such a
task, if not impossible, is highly nontrivial and has never been
carried out so far. For the latter problem relativity principles
require that the evolution of many-body systems, such as approaching
equilibrium must be observer-independent. Furthermore, it has been a
common belief that there exists covariant notion of evolution for
many-body systems \cite{Hakim67,Schieve89,Israel84,Kandrup84}. Yet,
it remains unclear how to reconcile these concepts with old
relativistic classical correlation dynamics.

A complete different formalism, but equally at the full microscopic
level, was formulated by Israel and Kandrup
\cite{Israel84,Kandrup84} in the framework of classical general
relativity (which is of minor importance). The Israel-Kandrup
formalism, although also starts from deterministic relativistic
classical dynamics of full interacting many-body systems, differs
from the old relativistic classical correlation dynamics in several
key aspects. First of all, it is manifestly covariant. There,
Hakim's analysis of relativistic statistical mechanics of $\mu$
(namely single particle) phase space is substantially extended to
$\Gamma$ (namely many particle) phase space. Then, the underlying
relativistic classical many-body dynamics is formulated in terms of
the so-called {\it action-at-a-distance formalism} (see, e.g.,
Ref.~\cite{Martinez06} for a review) rather than the Hamiltonian
formalism. Remarkably, in such a formalism fields are not considered
as independent degrees of freedom. Instead, they are carried by
complicated retarded (advanced) potentials. As a result, particles
interact in a nonlocal manner. And the force acting on given
particles must be viewed not only as a function of particle's
$4$-position and $4$-momentum vector, but also as a functional of
the world line of all the other particles. Finally, particles are
treated on different footing in the way that the $N$-particle system
is divided into $1$- and $(N-1)$-particle subsystem coupling to each
other. As such they managed to employ Willis-Picard projection
operator approach \cite{Willis74} to obtain a closed nonlinear
kinetic equation of one-body distribution function. The latter gives
the relativistic Landau equation
\cite{Belyaev57,Klimontovich60,Klimontovich67} from which, {\it at
the full microscopic level},  J{\"u}ttner equilibrium distribution
immediately follows. The success of the Israel-Kandrup formalism
justified a number of far-reaching concepts. Among of them are: the
global covariant evolution as conjectured by Hakim \cite{Hakim67}
and the legitimation of building relativistic classical
non-equilibrium statistical mechanics on the action-at-a-distance
formalism of deterministic many-body dynamics. The latter nowadays
is widely accepted \cite{Schieve89,Lu94}. Importantly, it was
suggested that the one-body distribution function may, at least
asymptotically, determine the covariant evolution of the system--the
very nature of the Bogoliubov functional assumption.

\subsection{Overview of this series of papers}
\label{outline}

Despite of the significant achievements by Israel and Kandrup there
are many important problems remained unsolved. First of all, it is
not clear how to go beyond the weak coupling approximation within
the Israel-Kandrup formalism. This issue is of great practical
importance especially for transport processes in plasmas with
electromagnetic interactions \cite{Klimontovich67,Lu94}. Also,
because of such a drawback one fails to predict--at the full
microscopic level--kinetic equations and (local) equilibrium for
dilute systems with moderate interaction strength. The solution to
this problem may solve the puzzle of the relativistic generalization
of the Maxwell-Boltzmann distribution which is currently undergoing
intensified debate \cite{Schieve89,Hanggi07,Debbasch08}. Then, in
Ref.~\cite{Kandrup84} the (local) equilibrium was studied under the
kinetic approximation. There, the principal problem of formulating
many-body equilibrium \cite{Hakim67} has not been attacked. Finally,
how to extend the Bogoliubov functional assumption rigorously
remains largely unexplored.

In this series of papers we substantially extend the investigations
of Israel and Kandrup \cite{Israel84,Kandrup84} and widely explore
these issues in the framework of special relativity. We present a
new, manifestly covariant classical correlation dynamics. As the
basic viewpoint we proceed along the line of
Refs.~\cite{Israel84,Kandrup84} and formulate the underlying
deterministic classical many-body dynamics in the
action-at-a-distance formalism with $N$ motion equations as
fundamental objects \cite{Fokker29,Wheeler49,Martinez06}. Such
formalism allows us to naturally work in physical coordinate and
momentum. The apparent advantage is to admit a theory which is
manifestly covariant at each step of the manipulations. Nevertheless
this is by no means merely based on the aesthetic viewpoint. Indeed,
we have been compelled to do so by significant progress achieved
recently by both physicists and mathematicians. (i) In a
$1$-dimensional numerical simulation it is found that point-like
collisions tend to drive a relativistic system into equilibrium
described by J{\"u}ttner distribution \cite{Hanggi07}. At the
microscopic level the deterministic classical many-body dynamics of
the underlying system may be described by the action-at-a-distance
formalism in a rather straightforward manner. A statistical theory
based on this formalism, in turn, is expected to explain the
experimental discovery. (ii) The Kirpichev-Polyakov theorem
\cite{Kirpichev07} partly justifies the longstanding conjecture of
Hakim on mathematical foundations of relativistic statistical
mechanics \cite{Hakim67}. It is shown that for $1$-dimensional
relativistic dynamics of classical charged systems, which is
formulated in terms of the Wheeler-Feynmann formalism
\cite{Wheeler49}, an ordinary Cauchy problem as in the Newtonian
mechanics can be stated. This is conceptually important because it
suggests that the global evolution of relativistic many-body systems
may be formulated in the way analogous to the Newtonian physics,
despite that
particles interact in a dramatically different manner.

The $N$ motion equations define a natural solution space namely the
$8N$-dimensional $\Gamma$ phase space. (Throughout this work the
mass-shell constraint is absorbed into the distribution functions.)
For this phase space we may define a distribution function which,
remarkably, depends on $N$ proper times. A significant difference
from the Newtonian physics is that a bundle of $N$ world lines,
rather than a representation point in the $\Gamma$ phase space
underlies the subsequent analysis of the dynamics of distribution
functions. Then, we formulate $N$
conservation equations. Moreover, for a large class of physical
forces such as Lorentz forces the phase space volume element is
invariant along world lines. As a result, despite of the absence of
Hamiltonian these conservation equations collapse into the
manifestly covariant Liouville equations. With the $N$ proper times
identified we obtain a single-time Liouville equation which
describes the manifestly covariant global evolution of distribution
function. On this global evolution we may build a manifestly
covariant theory of classical correlation dynamics by using the
Balescu-Wallenborn projection operator approach
\cite{Balescu,Balescu71}. It then follows that the reduced one-body
distribution function fully determines the $N$-body distribution
function, in particular, the entire correlation functions. The
evolution of the reduced one-body distribution function obeys an
exact closed kinetic equation. As such, we achieve the relativistic
Bogoliubov functional assumption which, in contrast to earlier
theories \cite{Naumov81,Lu94,Tian01}, is manifestly covariant. From
the exact closed kinetic equation we recover various (Vlasov,
Landau, Boltzmann) relativistic kinetic equations systematically.
The solutions to these kinetic equations allow us to pass to
macroscopic physical observables by carrying out appropriate average
with respect to them.

It must be stressed that the present theory cannot serve as an
alternative to the difficult problem of deterministic relativistic
classical many-body dynamics. Neither are they equivalent. As the
Newtonian physics in order to develop a theory of manifestly
covariant classical correlation dynamics one actually requires only
few assumptions regarding dynamical properties of deterministic
relativistic many-body systems. These assumptions are exactly
formulated in this paper. Although to prove them may be hard
mathematical problems, they have apparent physical implications.
Thus, these assumptions and thereby the present relativistic
classical correlation dynamics are expected to be applicable for a
large class of realistic relativistic many-body systems. It is also
worth pointing out that compared to its Newtonian counterpart the
classical correlation dynamics presented here suffers from
additional technical complications. That is, the force acting on
given particle is determined by the world lines of all the other
particles and, therefore, as one passes from the exact closed
kinetic equation to special kinetic equations appropriate
approximations regarding the world line must be implemented. Indeed,
in the present work we use the well known relativistic impulse
approximation \cite{Israel84,Kandrup84}. There, to fully determine
interactions between two particles at given moment their
trajectories are considered to be linear. To overcome this technical
complication one, in principle, needs to either expand phase space
so as to accommodate particle acceleration as well as its
higher-order derivatives \cite{Hakim67}, or treat fields as
independent degrees of freedom which is a tractable task. (However,
the $1$-dimensional relativistic dynamics of classical charged
systems may be an exception, because according to the
Kirpichev-Polyakov theorem the field degrees of freedom are
redundant.) Since the present work aims at the principle problem of
building a theory of manifestly covariant classical correlation
dynamics on the action-at-a-distance formalism, we may leave this
technical issue at this stage of conceptual development, in
particular, if we ignore problems such as radiation reaction (for
electromagnetic interactions).

We plan to explore various topics discussed above in this and the
following \cite{Tian09} paper. This first one is devoted to the
general mathematical structure and is organized as follows: In
Sec.~\ref{Liouville} we first introduce preliminary concepts
required for developing a theory of manifestly covariant global
evolution. Then we derive a manifestly covariant single-time
Liouville equation. In Sec.~\ref{subdynamics} the correlation
pattern representation is established in a manifestly covariant
manner. On this basis we apply the Balescu-Wallenborn projection
operator approach to the single-time Liouville equation. In
Sec.~\ref{Bogoliubovhypothesis} we present the relativistic version
of the Bogoliubov functional assumption. In particular, we prove the
factorization theorem and derive the exact closed kinetic equation
of reduced one-body distribution function. We close this paper by
outlining further applications. In particular, we present an
extensive discussion of generalizing the manifestly covariant
classical correlation dynamics to the general relativity so as to
treat self-gravitating systems, which is the main subject of
Sec.~\ref{remark}. Some technical details are presented in Appendix
\ref{BBGKY2} and \ref{partitionLiouville}.

The second paper (denoted as Paper II), written in a self-contained
manner, is devoted to applications of the general principles to
relativistic plasmas with electromagnetic interactions. The readers
interested only in applications of the present theory may skip this
first one and read the second one directly. There, the relativistic
Vlasov, Landau and Boltzmann equation follow systematically from the
exact manifestly covariant nonlinear equation which is closed, and
satisfied by the reduced one-body distribution function. The
collision integrals of the latter two justify the J{\"u}ttner
distribution as the special relativistic generalization of the
Maxwell-Boltzmann distribution. The collective effects, such as the
issue of correlation at the relativistic many-body equilibrium are
explored. We summarize this series of work in Paper II.

Finally we list some of the notations and conventions. We choose the
unit system with the speed of light $c=1$\,. To distinguish from the
Minkowski $4$-vector we use the bold font to denote the vector in
the Euclidean space. Greek indices running from $0$ to $3$ are
further used to denote the component of the former. The Einstein
summation convention is applied to these indices. The
$4$-dimensional Minkowski space is endowed with the metric
$\eta^{\mu\nu}={\rm diag}(1,-1,-1,-1)$\,. The scalar product of two
$4$-vectors is defined as $a\cdot b \equiv \eta^{\mu\nu} a_\mu b_\nu
= a_\mu b^\mu$\,. In particular, $a\cdot a\equiv a^2$\,. The
arguments of distribution functions: $(x_i,p_i,\tau_i+\tau)$ carried
by particle $i$ are abbreviated as $i$\,. In addition to the usual
mathematical symbols we use the following notations:
\\
\\
\begin{tabular*}{0.30\textwidth}{@{\extracolsep{\fill}}  l l   }
$\partial_\mu$\,, & covariant derivative: \, $\partial_\mu =
\partial/\partial x^\mu$ \,; \\
$d^4 z$\,, & volume element in $4$-dimensional Minkowski \\
& space: \, $d^4 z= dz^0 dz^1 dz^2 dz^3$\,;\\
$\delta^{(d)}(f)$\,, & $d$-dimensional
Dirac function:\\
$d\Sigma_\mu$\,, & differential form of spacelike $3$-surface:\\
& $d\Sigma_\mu = \frac{1}{3!} \epsilon_{\mu\nu\rho\lambda} \, dx^\nu
\wedge dx^\rho \wedge dx^\lambda $ with
$\epsilon_{\mu\nu\rho\lambda}$ \\
& being $ \pm 1$ when
$(\mu\nu\rho\lambda)$ is an even (odd) \\
& permutation of $(0123)$ and
being $0$ otherwise\,;\\
$x_i[\varsigma]$\,, & world line of particle $i$\,;\\
$x_i(\varsigma)$\,, & $4$-position of particle $i$ at proper time $\varsigma$\,;\\
$ di$\,, & volume element in the $\mu$ phase space of \\
 & particle $i$\,:\, $di = d^4 x_i d^4 p_i $\,.
\end{tabular*}

\section{Manifestly covariant Liouville equations}
\label{Liouville}

Let us consider a system consisting of $N$ classical point
particles, each of which has the rest mass $m$\,. The particles have
the action at a distance on other particles. The interaction
propagates at the speed of light. (To simplify discussions
throughout this series of papers we shall not consider the
self-action.) The interactions carry energy and momentum to and from
particles without the support of independent fields. This is the
so-called action-at-a-distance formalism of relativistic classical
many-body dynamics pioneered by Fokker \cite{Fokker29} and Wheeler
and Feynman \cite{Wheeler49}. The subject has been widely explored
in the last half century \cite{Martinez06}. In this section based on
this formalism we derive a manifestly covariant single-time
Liouville equation. Analogous to the Newtonian physics it then
serves as the exact starting point of relativistic classical
correlation dynamics.

\subsection{Action-at-a-distance formalism}
\label{mechanics}

The history of the microscopic system is described by a bundle of
$N$ particle world lines which solve the following relativistic
motion equations:
\begin{eqnarray}
\frac{dx^\mu_i}{d\tau_i} &=& \frac{p^\mu_i}{m}\equiv
u^\mu_i\,,
\label{Newton1}\\
\frac{dp^\mu_i}{d\tau_i} &=& F^\mu_{\rm
ext}(x_i,p_i)+\sum_{j\neq i}^N
F^\mu_{ij}(x_i,p_i)\,. \label{Newton}
\end{eqnarray}
Here $x_i^\mu(\tau_i)\,, u_i^\mu(\tau_i)\,, p_i^\mu(\tau_i)$ are the
$4$-position, $4$-velocity and $4$-momentum vector of particle $i$
depending on the proper time $\tau_i$\,, respectively, $F^\mu_{\rm
ext}$ is the the external force, and $F^\mu_{ij}$ is the force
acting on particle $i$ by particle $j$\,. Importantly, we assume
that both forces do not depend on the acceleration of the acted
particle, and consider the interacting force $F_{ij}^\mu$ with the
general form as follows \cite{Martinez06}:
\begin{eqnarray}
&&
F_{ij}^\mu(x_i,p_i)
\label{force} \\
&=& \int_{-\infty}^{+\infty}d\tau_j\, s(\rho_{ij})\, {\cal F}^{\mu
\nu}(\alpha_{ij},\beta_{ij},\gamma_{ij},\gamma_{ji},\zeta_{ij})
\, p_{\nu i} \,, \nonumber
\end{eqnarray}
where ${\cal F}^{\mu \nu}$ is an antisymmetric tensor, and the role
of function $s (\rho_{ij})$ is to invariantly connect $x_i$ with one
(or several) points at the world line $x_j(\tau_j)$\,. The arguments
in Eq.~(\ref{force}) are defined as follows:
\begin{eqnarray}
&& \qquad \qquad  \rho_{ij} = (x_{\mu i}-x_{\mu j}) (x^\mu_i-x^\mu_j)\,,\nonumber\\
&& \qquad \qquad  \alpha_{ij}=(x^\mu_i-x^\mu_j) p^\nu_j-p^\mu_j (x^\nu_i-x^\nu_j)\,,
\nonumber\\
&& \beta_{ij}=\frac{dp_j^\mu}{d\tau_j}\, (x_{\mu i}-x_{\mu
j})\,, \gamma_{ij}= p_i^\mu (x_{\mu j}-x_{\mu i})
\,, \nonumber\\
&& \qquad \zeta_{ij}= (x^\mu_i-x^\mu_j)
\frac{dp^\nu_j}{d\tau_j}-\frac{dp^\mu_j}{d\tau_j} (x^\nu_i-x^\nu_j)
\,. \label{force1}
\end{eqnarray}
The relation: $\qquad p_i\cdot F_{ij}(x_i,p_i)=0$
is obvious because of the skew property of ${\cal F}^{\mu \nu}$\,.
Such a relation is further enforced to external forces:
\begin{equation}
p_i\cdot F_{\rm ext}(x_i,p_i)=0 \,. \label{forcecondition2}
\end{equation}
Therefore, the mass-shell constraint:
\begin{equation}
p_i^2=m^2 \label{forcecondition3}
\end{equation}
is guaranteed. To fully determine the interacting
force $F^\mu_{ij}$ relies on the world line of particle $j$ namely
$x_j(\tau_j)$\,.
In Paper II we exemplify the interacting force in the case
of electromagnetic interactions.

\subsection{Manifestly covariant single-time Liouville equation}
\label{statics}

The motion equations (\ref{Newton1}) and (\ref{Newton}) suggest
that we may introduce the concept of
$8N$-dimensional $\Gamma$ phase space. It is defined as the direct
product of $N$ flat $\mu$ phase space, each of which corresponds to
individual particle. That is,
\begin{eqnarray}
\Gamma = \bigotimes_{i=1}^N\, \mu_i\,, \qquad \mu_i = {\cal M}_i^4
\otimes U_i^4\,,
\label{space}
\end{eqnarray}
where ${\cal M}_i^4$ and $U_i^4$ are the $4$-dimensional Minkowski
spacetime and momentum space carried by particle $i$\,,
respectively. The volume element of the $\Gamma$ phase space is
$d\Gamma=\prod_{i=1}^N\,di\,, di\equiv d^4 x_i d^4 p_i $\,. There
exists a fundamental difference between the $\Gamma$ phase space
defined above and its Newtonian counterpart: In contrast to the
Newtonian physics $X_N\equiv (x_1(\tau_1),p_1(\tau_1),\cdots\,,
x_N(\tau_N),p_N(\tau_N))\in \Gamma$ generally does not represent a
microscopic state because, in general, it is not sufficient to
uniquely determine the history of the full system namely the $N$
particle world lines. Instead, in special relativity microscopic
states have been conjectured to be $N$ segments of particle world
lines each of which is carried by individual particle
\cite{Wheeler49}. Furthermore, all the microscopic states constitute
the space $\Omega^N$ which has the dimensionality $\geq 8N$\,.
(Ref.~\cite{Kirpichev07} presents a nontrivial example where the
equality holds.)

Given a bundle of $N$ particle world lines satisfying
Eq.~(\ref{Newton}), or more precisely, a microscopic state
${\tilde \omega}\in \Omega^N$ the representation point $X_N$ has the singular
distribution as $\prod_{i=1}^N\,
\delta^{(4)}(x_i-x_i(\tau_i))\delta^{(4)}(p_i-p_i(\tau_i))$\,, which
depends on ${\tilde \omega}$ through the phase trajectory
$x_i(\tau_i),p_i(\tau_i)\,, i=1,\cdots\,, N$\,. By preparing a
cloud of microscopic states centering at ${\tilde \omega}$ and carrying out the
average over this ensemble, we expect to smear the singular
distribution at least along some directions (of the $\Gamma$ phase space). More
precisely, we wish to assert the
existence of the following \\
\\
\noindent Assumption 2.1 {\it There exists a probability measure
on $\Omega^N$ such that the average below
\begin{eqnarray}
&& {\cal D}(x_1,p_1,\tau_1, \cdots
\,,x_N,p_N,\tau_N;x_{1}[\varsigma],\cdots,x_{N}[\varsigma]) \nonumber\\
&\equiv&
\left\langle \prod_{i=1}^N\, \delta^{(4)}(x_i-x_i(\tau_i))\,
\delta^{(4)}(p_i-p_i(\tau_i))\right\rangle
\label{density}
\end{eqnarray}
strongly peaks at $(x_1(\tau_1), p_1(\tau_1),\cdots\,, x_N(\tau_N), p_N(\tau_N))$\,.}\\

{\it Remark.} Although the details of the average are not required for
subsequent discussions, it is important to notice that ${\cal D}$ possesses
$N$ proper times and functionally depends on
the $N$ particle world lines, i.e., $x_{1}[\varsigma],\cdots,x_{N}[\varsigma]$\,.
Physically, it may be interpreted as a
probability distribution function \cite{note} in the $\Gamma$ phase space. Indeed, it is
obvious that ${\cal D}$ has the following properties: (i) ${\cal
D}\geq 0$ and (ii) $\int d\Gamma\, {\cal D}=1$\,. Given
an elementary Lebesgue-measurable subset $\omega=\otimes_{i=1}^N\,
\omega_i \subset \Gamma $\,, $\int_\omega d\Gamma \, {\cal D}$
stands for the probability for particle $i$ to be in $\omega_i$ in
$\tau_i$ ($i=1,\cdots\,, N$)\,.

Similar to relativistic nonequilibrium mechanics built on the $\mu$
phase space \cite{Hakim67}, from the causality follows\\
\\
\noindent Lemma 2.2. {\it The distribution function ${\cal D}$ satisfies
\begin{equation}
\lim_{\tau_i\rightarrow \pm \infty} {\cal D}=0\,, \quad \forall\, i
\label{causality}
\end{equation}
in the sense of Lebesgue measure.}\\
\\

{\it Proof.} Consider the probability density at the proper times
$(\tau_1,\cdots\,, \tau_N)$ defined as
\begin{eqnarray}
\rho (x_1,\tau_1, \cdots
\,,x_N,\tau_N;x_{1}[\varsigma],\cdots,x_{N}[\varsigma])=
\int\!\! \prod_{i=1}^N d^4 p_i{\cal D}.
\label{causality1}
\end{eqnarray}
Then, the probability for particle $i$ to stay inside a finite Lebesgue-measurable volume
${\bar \omega}_i\in {\cal M}^4_i$ at the proper time $\tau_i$ ($i=1,\cdots\,, N$) is
\begin{eqnarray}
n(\tau_1,\cdots\,, \tau_N) = \int_{{\bar \omega}_1}d^4x_1\cdots \int_{{\bar \omega}_N}d^4x_N \, \rho \,.
\label{causality2}
\end{eqnarray}
Notice that for given $i$ the world line $x_i[\varsigma]$\,, having crossed ${\bar \omega}_i$\,,
could not return. Thus,
\begin{eqnarray}
\lim_{\tau_i\rightarrow \pm \infty}
n(\tau_1,\cdots\,, \tau_N) = 0
\label{causality3}
\end{eqnarray}
for arbitrary ${\bar \omega}_i$\,. From this the lemma follows.
Q.E.D.\\

However, the distribution function ${\cal D}$ has a remarkable
property. That is, with the momenta integrated out it is reduced
into a distribution function defined on the entire Minkowski spacetime. Thus, the
distribution function ${\cal D}$ is not physical. Instead, to match
macroscopic observations we have to further introduce the physical
distribution function ${\cal N}$ such that with momenta
integrated out it is reduced into a distribution function defined on a spacelike
$3$-surface--a subspace of the Minkowski spacetime. Mathematically, it is
defined as
\begin{eqnarray}
{\cal N}(x_1,p_1,\cdots
\,,x_N,p_N;x_{1}[\varsigma],\cdots,x_{N}[\varsigma])
\! \equiv \! \int
\!\!
\prod_{i=1}^N d\tau_i
{\cal D}
\,,
\label{density1}
\end{eqnarray}
which is the straightforward generalization of its $\mu$ phase
space counterpart \cite{Hakim67}. The normalization of ${\cal N}$ follows from
\\
\\
\noindent Lemma 2.3. {\it The physical distribution function ${\cal
N}$ satisfies
\begin{eqnarray}
\int_{\Sigma_1\otimes U_1^4} d\Sigma_{\mu_1} d^4 p_1 u_1^\mu \cdots
\int_{\Sigma_N\otimes U_N^4} d\Sigma_{\mu_N} d^4 p_N u_N^\mu\, {\cal
N} =1 \,.
\nonumber\\
\label{normalization2}
\end{eqnarray}
Here $\Sigma_i$ is an arbitrary spacelike $3$-surface in ${\cal
M}_i^4$ with the vectorial value of the differential form as $d\Sigma_{\mu_i}$\,.}\\

{\it Proof.} 
For the volume element $d^4x_i$ we have
\begin{equation}
d^4 x_i = d\Sigma_{\mu_i} dx^\mu_i = d\tau_i d\Sigma_{\mu_i}
\frac{dx^\mu_i}{d\tau_i}
\,,
\label{volumeelement}
\end{equation}
where $d\Sigma_{\mu i}$ is the differential form of an arbitrary
spacelike $3$-surface $\Sigma_i$\,:
\begin{equation}
d\Sigma_{\mu i} = \frac{1}{3!} \epsilon_{\mu\nu\rho\lambda} \,
dx_i^\nu \wedge dx_i^\rho \wedge dx_i^\lambda \,.
\label{differential}
\end{equation}
Here $\epsilon_{\mu\nu\rho\lambda}$ is $ \pm 1$ when
$(\mu\nu\rho\lambda)$ is an even (odd) permutation of $(0123)$ and
is $0$ otherwise.
Then,
\begin{eqnarray}
&& d^4 x_i d^4 p_i\delta^{(4)}(x_i-x_i(\tau_i)) \delta^{(4)}(p_i-p_i(\tau_i))\label{normalization5}\\
&=&
 d\tau_i d\Sigma_{\mu_i} \frac{dx^\mu_i}{d\tau_i} d^4 p_i
 \delta^{(4)}(x_i-x_i(\tau_i)) \delta^{(4)}(p_i-p_i(\tau_i)) \nonumber\\
&=& d\tau_i d\Sigma_{\mu_i} d^4 p_i\, u^\mu_i
 \delta^{(4)}(x_i-x_i(\tau_i)) \delta^{(4)}(p_i-p_i(\tau_i))\,.
\nonumber
\end{eqnarray}
Notice that $\int d\Gamma\, {\cal D}=1$\,. We then substitute
Eq.~(\ref{normalization5}) into it and integrate out $\tau_i$'s. With
Eqs.~(\ref{density}) and (\ref{density1}) taken into account eventually we obtain
Eq.~(\ref{normalization2}). Q.E.D.\\

{\it Remark.} This lemma shows that the physical distribution
function ${\cal N}$ is the analogue of the $\Gamma$ phase space
distribution function in the Newtonian physics. The $7N$-dimensional
manifold on which ${\cal N}$ is normalized may be considered as the
effective phase space. For the convenience below here we also
introduce the space of physical one-body distribution ${\cal H}$
defined as:
\begin{eqnarray}
{\cal H} &\equiv& \bigg \{h| h: \Sigma\otimes U^4 \mapsto \mathbb{R}^+
\cup \{0\}\,, \nonumber\\
&& \lim_{N\rightarrow +\infty }N^{-1}\int_{\Sigma\otimes
U^4} d\Sigma_{\mu} d^4 p\, u^\mu\, h=1 \bigg\} \,.
\label{physicaldistributionspace}
\end{eqnarray}

To further proceed we introduce the Liouvillian $\Li$ which is
decomposed into the free and interacting part, i.e., ${\hat
{\mathfrak L}}={\hat {\mathfrak L}}^0 + \lambda {\hat {\mathfrak
L}}' $ with ${\hat {\mathfrak L}}^0$ and $\lambda {\hat {\mathfrak
L}}'$ the free and interacting Liouvillian, respectively. They are
defined as
\begin{eqnarray}
{\hat {\mathfrak L}}^0 &=& -\sum_{i=1}^N \,\left[ u_i^\mu
\partial_{\mu i} + F^\mu_{\rm ext}
(x_i,p_i)\frac{\partial}{\partial p_i^\mu}\right]\,, \nonumber\\
\lambda {\hat {\mathfrak L}}' &=& \sum_{i<j}\, \lambda
\LI'_{ij}\,, \nonumber\\
\lambda \LI'_{ij} &\equiv& - \left\{F^\mu_{ij}
(x_i,p_i)\frac{\partial}{\partial p_i^\mu} + F^\mu_{ji}
(x_j,p_j)\frac{\partial}{\partial p_j^\mu} \right\}\,,
\label{interactingLiouville}
\end{eqnarray}
where $\lambda \LI'_{ij}$ is the two-body interacting Liouvillian.
Notice that here the dimensionless parameter $\lambda$ characterizes
the interaction strength. Now we are ready to prove the following
Liouville theorem which justifies the covariant notion of evolution in the $\Gamma$ phase space,
and constitutes the exact starting point of the succeeding
sections: \\

\noindent Theorem 2.4. {\it If both the external and interaction
force are conservative, i.e.,
\begin{equation}
\frac{\partial}{\partial p_i^\mu} F^\mu_{\rm ext}(x_i,p_i)=0\,,
\qquad \frac{\partial}{\partial p_i^\mu} F^\mu_{ij}(x_i,p_i)=0\,,
\label{forcecondition}
\end{equation}
then the distribution function ${\cal D}
(x_1,p_1,\tau_1+\tau,\cdots \,,x_N,p_N,\tau_N+\tau;x_{1}[\varsigma],\cdots,x_{N}[\varsigma])$ satisfies the following
Liouville equation:
\begin{eqnarray}
\left (\frac{\partial}{\partial \tau}- {\hat {\mathfrak L}}\right)
{\cal D}
=0 \,.
\label{Liouville1}
\end{eqnarray}
}\\

{\it Proof.} Suppose that at the proper times $(\varsigma_1,\cdots\,,
\varsigma_N)$ particles are in the volume $\omega=\prod_{i=1}^N\,
d^4 x_i(\varsigma_i)d^4 p_i(\varsigma_i)$ with a probability density
${\cal D}
(x_1,p_1,\tau_1+\tau,\cdots \,,x_N,p_N,\tau_N+\tau;x_{1}[\varsigma],\cdots,x_{N}[\varsigma])$\,.
Now let some particle
say $i$ evolves from $\varsigma_i$ to $\varsigma'_i$ along the world
line $x_i[\varsigma]$\,, while others frozen. As such the
probability is conserved. That is,
\begin{eqnarray}
&& d^4 x_i(\varsigma_i) d^4 p_i(\varsigma_i)
\prod_{j\neq i} d^4 x_j d^4 p_j
 {\cal D}(\cdots x_i(\varsigma_i),p_i(\varsigma_i),\varsigma_i
\cdots )
\nonumber\\
&=& d^4 x_i(\varsigma'_i) d^4 p_i(\varsigma'_i)
\prod_{j\neq i} d^4 x_j d^4 p_j
{\cal D}(\cdots
x_i(\varsigma'_i),p_i(\varsigma'_i),\varsigma'_i\cdots)
\nonumber\\
\label{conservation}
\end{eqnarray}
with all the irrelevant arguments in ${\cal D}$ suppressed.
Since the forces are
conservative, the Lebesgue measure is conserved, i.e., $d^4x_i
(\varsigma_i)d^4p_i (\varsigma_i)=d^4x_i (\varsigma'_i)d^4p_i
(\varsigma'_i)$\,. Eq.~(\ref{conservation}) then gives
\begin{eqnarray}
{\cal D}(\cdots x_i(\varsigma_i),p_i(\varsigma_i), \varsigma_i
\cdots ) = {\cal D}(\cdots
x_i(\varsigma'_i),p_i(\varsigma'_i),\varsigma'_i\cdots) \,.
\nonumber\\
\label{densityinvariance}
\end{eqnarray}
Letting $\varsigma'_i \rightarrow \varsigma_i$ we obtain
\begin{eqnarray}
\left \{\frac{\partial}{\partial \varsigma_i}+ u_i^\mu
\partial_{\mu i} + \left[F^\mu_{\rm ext}(x_i,p_i)+\sum_{j\neq i}^N F^\mu_{ij}(x_i,p_i)\right]
\frac{\partial}{\partial p_i^\mu}\right \}  \nonumber\\
{\cal
D}(x_1,p_1,\varsigma_1,\cdots \,,x_N,p_N,\varsigma_N;x_{1}[\varsigma],\cdots,x_{N}[\varsigma])
= 0 \,.\nonumber\\
\label{Liouvillepartial}
\end{eqnarray}
Totally we have $N$ such Liouville equations with different
evolution time $\varsigma_i$\,.

Furthermore, we let the $N$ proper times change uniformly, i.e.,
\begin{equation}
d\varsigma_1 = \cdots =d\varsigma_N \equiv d\tau
\label{time}
\end{equation}
or equivalently, $\varsigma_i =\tau_i +\tau$ with $\tau_i$'s the
initial proper times. Then, adding the $N$ Liouville equations
together
Eq.~(\ref{Liouville1}) follows. Q.E.D.\\

{\it Remarks.} (i) In the absence of interactions the above
single-time Liouville equation was derived by Hakim \cite{Hakim67}.
The manifestly covariant Liouville equation originates at the
probability conservation law, and is irrespective of the Hamiltonian
character of relativistic classical many-body dynamics. Of course,
in the Hamiltonian formalism the (single-time) Liouville equation
trivially exists albeit nonmanifestly covariant
\cite{Prigogine65,Balescu67}. In the action-at-a-distance formalism
the classical many-body dynamics has non-Hamiltonian character. To
justify the Liouville equation the constraint of conservative force
must be imposed. Provided that such constraint is released one may
enlarge the $\Gamma$ phase space and subsequently arrive at a
generalized single-time Liouville equation \cite{Hakim67}.

(ii) The single-time Liouville equation shows that the distribution
function ${\cal D}$ depends on a proper time $\tau$ parametrizing
the global evolution in the $\Gamma$ phase space. Yet, it still
depends on the other $N-1$ proper times associated with the initial
condition of the global evolution. Upon passing from $N$ Liouville
equations (\ref{Liouvillepartial}) to the single-time Liouville
equation (\ref{Liouville1}) some details of deterministic classical
many-body dynamics are lost. Thus, the single-time Liouville
equation is not equivalent to relativistic motion equations. The
peculiar feature of multiple proper times roots in the
noninstantaneous feature of relativistic force and causes a
conceptual difference from the Newtonian physics. In the latter case
the instantaneous nature of forces allows one to parametrize the
global evolution by the usual coordinate time. Consequently, in the
sense of dynamics of distribution functions (single-time) Liouville
equation is equivalent to motion equations.

(iii)
For Eq.~(\ref{Liouville1}) let us integrate out
$\tau_i$'s. Taking into account
Eq.~(\ref{density1}) we find
\begin{eqnarray}
\Li {\cal N} (x_1,p_1,\cdots
\,,x_N,p_N;x_{1}[\varsigma],\cdots,x_{N}[\varsigma])=0 \,.
\label{Liouville2}
\end{eqnarray}
Provided that the functional dependence of ${\cal N}$ on
$x_{1}[\varsigma],\cdots,x_{N}[\varsigma]$ is released it then
recovers the exact starting point of Refs.~\cite{Israel84,Kandrup84}
(in the framework of general relativity) and Ref.~\cite{Lu94}. The
very nature of such functional independence on the $N$ particle
world lines is that at given global proper time $\tau$ the
representation point is sufficient to uniquely determine the
particle world lines. This does happen in the asymptotic sense. It
turns out that, as the Newtonian physics, for sufficiently large
proper time $\tau$ the relativistic many-body systems may also lose
the memory on ``initial'' world line segments--the profound change
of deterministic classical dynamics. Consequently, the instantaneous
evolution of ${\cal N}$ is determined by specified phase
trajectories associated with $N$ particles each of which,
importantly, merely depends on the phase coordinate of given
particle. In Sec.~\ref{Bogoliubovhypothesis} we will show that the
single-time Liouville equation (\ref{Liouville1}), indeed, admits
such kind of solution in the thermodynamic limit: $N\rightarrow
+\infty$\,. However, we are not able to justify this picture for
finite but sufficiently large $N$\,.

\section{Correlation dynamics analysis of global evolution}
\label{subdynamics}

In this section we wish to apply the Balescu-Wallenborn projection
operator approach \cite{Balescu71,Balescu} to the single-time
Liouville equation (\ref{Liouville1}). In doing so we achieve a
theory of relativistic classical correlation dynamics in the way
that at each step the manipulations are manifestly covariant. With
this approach we manage to split the proper-time dependent
distribution function into the kinetic and non-kinetic component.
Each of them independently evolves, and the latter decays for
sufficiently large global proper time.

It should be emphasized that given appropriate assumptions regarding
the deterministic classical many-body dynamics, as to be specified
below, (In other words, we will deal with some {\it axiom dynamical
system}.) all the results presented in Sec.~\ref{subdynamics} and
\ref{Bogoliubovhypothesis} are mathematically rigorous.

\subsection{Reduced distribution function representation}
\label{reduceddistribution}

Let us first introduce the concept of reduced distribution function.
From now on we denote the arguments:
$(x_i,p_i,\tau_i +\tau)$ of distribution functions as $i$\,.
Then, the reduced
$s$-body distribution function is defined as
\begin{eqnarray}
&& {\cal D}_s(i_1, \cdots
\,,i_s;x_{i_1}[\varsigma],\cdots,x_{i_s}[\varsigma]) \nonumber\\
&\equiv& \int
\prod_{j=s+1}^N di_j\, {\cal D}(1, \cdots
\,,N;x_{1}[\varsigma],\cdots,x_{N}[\varsigma]) \,, \nonumber\\
&& \qquad \qquad \qquad \forall \,
1\leq i_1<\cdots <i_s \leq N\,. \label{reducedDensity}
\end{eqnarray}
Notice that here the particle groups $(i_1, \cdots
\,,i_s)$ and $(i_{s+1}, \cdots
\,,i_N)$ constitute a partition of the full system $(1,\cdots\,, N)$\,. Moreover,
the reduced $s$-body
distribution function ${\cal D}_s$ stands for the probability density for
particle $i_j$ to be at $(x_{i_j}\,, p_{i_j})$ at the proper time
$\tau_{i_j}+\tau$ ($1\leq j\leq s$)\,. The normalization can be shown
to be
\begin{equation}
\int \prod_{j=1}^s di_j\, {\cal D}_s(i_1, \cdots
\,,i_s;x_{i_1}[\varsigma],\cdots,x_{i_s}[\varsigma]) =1 \,.
\label{normalization1}
\end{equation}

Then, for Eq.~(\ref{Liouville1}) we integrate out the phase
coordinates of particle $i_{s+1}, \cdots \,,i_N$\,. As a result we
obtain
\begin{eqnarray}
&& \left\{\frac{\partial}{\partial \tau} + \sum_{j=1}^s\,
u^\mu_{i_j}\partial_{\mu i_j} - \sum_{j< j'}^s \, \lambda
\LI'_{i_ji_{j'}}\right\}{\cal D}_s
\nonumber\\
&=& \sum_{{\small
\begin{array}{c}
                                             i_{s+1}=1 \\
                                             i_{s+1}\neq i_1,\cdots,i_s
                                           \end{array}
}}^N \int di_{s+1}\, \sum_{j=1}^s \, \lambda \LI'_{i_ji_{s+1}} {\cal
D}_{s+1}
\,,
\label{BBGKY}
\end{eqnarray}
which is a manifestly covariant relativistic BBGKY hierarchy. Here
in order to make the formula compact we have omitted all the
arguments of ${\cal D}_s$ and ${\cal D}_{s+1}$\,. Notice that
Eq.~(\ref{BBGKY}) differs from that studied in
Refs.~\cite{Naumov81,Lu94,Tian01}, where nonmanifestly covariant
relativistic BBGKY hierarchies were derived for physical
distribution functions.

Let us further define the distribution vector:
\begin{eqnarray}
\overrightarrow{{\frak D}} \equiv (\{{\cal D}_1\} \,, \{{\cal D}_2\}
\,, \cdots \,,\{{\cal D}_N \}\equiv {\cal D})
\,. \label{distributionvector}
\end{eqnarray}
Notice that,
for each $\{{\cal D}_s\}$ there are $N!/[(N-s!)s!]$ components each
of which corresponds to a $s$-particle group $(i_1,\cdots\,, i_s)$ with $1\leq i_1< \cdots <
i_s\leq N$\,. With the help of this definition the BBGKY hierarchy
is rewritten in a compact form:
\begin{equation}
\left (\frac{\partial}{\partial \tau}- {\hat {\mathfrak
L}}\right)\overrightarrow{{\frak D}}=0 \,.
\label{BBGKY1}
\end{equation}
The projection to general $s$-particle states, denoted as
$(i_1\,,\cdots\,, i_s|$ is identical to Eq.~(\ref{BBGKY}) provided
that the matrix elements are set to be
\begin{eqnarray}
&& (i_1\,,\cdots\,, i_j| \Li^0 |i_1\,,\cdots\,, i_j) = -\sum_{k=1}^j\,
u^\mu_{i_k}\partial_{\mu i_k}\,, \nonumber\\
&& (i_1\,,\cdots\,, i_j|
\lambda \Li' |i_1\,,\cdots\,, i_j,i_{j+1})  \nonumber\\
&& \qquad \qquad \qquad = -\int di_{j+1}\,
\sum_{k=1}^j \, \lambda \LI'_{i_ki_{j+1}}
\label{BBGKY3}
\end{eqnarray}
and to be zero otherwise. Notice that the two-body Liouvillian $\lambda
\LI'_{i_ki_{j+1}}$ is determined by the world lines of particles
$i_k$ and $i_{j+1}$\,.
Eqs.~(\ref{distributionvector})-(\ref{BBGKY3}) may be considered as
the {\it reduced distribution function representation} of the
single-time Liouville equation (\ref{Liouville1}).

Likewise, we may carry out the same program for the physical
distribution function ${\cal N}$ and cast Liouville equation
(\ref{Liouville2}) into another manifestly covariant BBGKY
hierarchy. The details are presented in Appendix~\ref{BBGKY2}.

\subsection{Correlation pattern representation}
\label{VC}

To proceed further we introduce the so-called {\it correlation
pattern representation}. Consider a general $s$-particle object
$F(1,\cdots,s;x_{k_1}[\varsigma],\cdots,x_{k_s}[\varsigma])$
possessing the permutation symmetry with respect to $12\cdots s$\,,
where $1\leq k_i\leq N\,,$ and $k_i \neq k_{i'}\,, \forall \,
i\neq i'$\,.
A unique correlation pattern, denoted as $|\Gamma_s\rangle$ (or
$\langle \Gamma_s|$) may be assigned to it in the following way: If
$F$ is factorized into $s$ components each of which merely depends
on $i\equiv (x_i,p_i,\tau_i+\tau)$ and $x_{k_i}[\varsigma]$ ($1\leq
i\leq s$), then $|\Gamma_s\rangle \equiv |0_s\rangle = |1|2|\cdots
|s\rangle $ (or $\langle \Gamma_s| = \langle 1|2|\cdots |s|$)\,,
which is called {\it vacuum state}
. In general, $F$ is at most partially factorized and called {\it
correlation state}. And the correlation pattern describes the
factorization structure. More precisely, suppose that $F$ possesses the
factorization as follows:
\begin{eqnarray}
&& F(1,\cdots,s;x_{k_1}[\varsigma],\cdots,x_{k_s}[\varsigma]) \nonumber\\
&=& \prod_{i=1}^j \, F_i({\rm
P}_i;x_{k_{s_i}}[\varsigma]\,,x_{k_{s_i+1}}[\varsigma]\,,\cdots\,,
x_{k_{s_i+j_i}}[\varsigma])\,, \label{vacuumstate}
\end{eqnarray}
where ${\rm P}_i\equiv
({s_i},{s_{i}+1},\cdots \,,{s_i+j_i})\neq \emptyset$\,, $i=1,2,\cdots\,,
j<N$ constitute a partition of $(1,2,\cdots \,,s)$\,:
\begin{eqnarray}
{\rm P}_1 \cup\cdots \cup {\rm P}_j = (1,\cdots \,,s),\,  {\rm
P}_i \cap {\rm P}_{i'}=\emptyset \,, \forall\, i\neq i' \,.
\label{correlationstate}
\end{eqnarray}
Then, the correlation pattern is $|\Gamma_s\rangle=|{\rm P_1}|{\rm
P_2}|\cdots |{\rm P_j}\rangle$ (or $\langle \Gamma_s|=\langle {\rm
P_1}|{\rm P_2}|\cdots |{\rm P_j}|$). Notice that the correlation
pattern possesses the permutation symmetry with respect to ${\rm
P_1}{\rm P_2}\cdots {\rm P_j}$\,. Therefore, two correlation patterns
are considered to be identical if they differ only in the order of ${\rm P}_i$\,.

With this definition the distribution vector may be decomposed in a
more delicate manner:
\begin{widetext}
\begin{eqnarray}
&& {\cal D}_s(i_1, \cdots
\,,i_s;x_{i_1}[\varsigma],\cdots,x_{i_s}[\varsigma]) \nonumber\\
&=& \sum_{\Gamma_s}\, |\Gamma_s\rangle \langle \Gamma_s|{\cal
D}_s(i_1, \cdots
\,,i_s;x_{i_1}[\varsigma],\cdots,x_{i_s}[\varsigma]) \nonumber\\
&=& \sum_{j=1}^s \sum_{{\rm P}_1\cdots {\rm P}_j}\, |{\rm P_1}|{\rm
P_2}|\cdots |{\rm P_j}\rangle \langle {\rm P_1}|{\rm P_2}|\cdots
|{\rm P_j}|{\cal D}_s(i_1, \cdots
\,,i_s;x_{i_1}[\varsigma],\cdots,x_{i_s}[\varsigma])
\label{partition}
\end{eqnarray}
\end{widetext}
\noindent
for general reduced $s$-body distribution function, which is the
reformulation of the cluster expansion. Here ${\rm P}_1\,,\cdots
\,,{\rm P}_j$ is a partition of $(i_1\cdots i_s)$\,. Notice that in the third
line any two terms do not possess identical correlation pattern. Furthermore, in
Appendix~\ref{partitionLiouville} we give the matrix elements of the
preliminary operators $\Li^0$ and $\lambda \Li'$ in this
representation. It should be stressed that such a cluster expansion
differs from traditional one \cite{Balescu} in that the distribution
functions depend on particle world lines. Such a concept was
first introduced by Klimontovich in the Newtonian context
\cite{Klimontovich60} and was generalized to special relativity--in a manifestly
covariant manner--by Hakim \cite{Hakim67}.

In this way we have assigned all the preliminary quantities to be
used below, i.e., $\overrightarrow{\mathfrak{D}}$\,, $\Li^0$ and
$\lambda \Li'$ a unique decomposition in the correlation pattern
representation. Thus, we may define the following vacuum and
correlation operator, denoted as $\V$ and $\C$\,, respectively:
\begin{eqnarray}
\V\, |\Gamma_r \rangle = \delta_{0_r\Gamma_r}\, |\Gamma_r \rangle
\,, \qquad \C\, |\Gamma_r \rangle = (1-\delta_{0_r\Gamma_r})\,
|\Gamma_r \rangle \label{VCdefinition}
\end{eqnarray}
which are diagonal in the correlation pattern
representation $|\Gamma_r\rangle$ (and also in $\langle \Gamma_r|$). It is easy to show that
$\V$ and $\C$ constitute an orthogonal decomposition, i.e.,
\begin{eqnarray}
\V+\C = {\bf 1}\,, \qquad \qquad \qquad \nonumber\\
\V^2 =\V\,, \C^2 = \C\,, \V\C=\C\V ={\bf 0} \,.
\label{VCrelation}
\end{eqnarray}
With the correlation pattern representations of
$\overrightarrow{\mathfrak{D}}\,, \Li^0\,, \lambda \Li'\,, \V$ and
$\C$ as building blocks, one may proceed to establish
representations of
more complicated operators with the help of following properties:
\begin{enumerate}
    \item {\it Completeness.}\, $\sum_{r}\sum_{\Gamma_r}\,
|\Gamma_r\rangle\langle \Gamma_r|={\bf 1}$\,;
    \item {\it Orthogonality.}\, $\langle
\Gamma_r|\Gamma'_{r'}\rangle=\delta_{rr'}\, \delta_{\Gamma_r
\Gamma'_{r'}}$\,;
    \item {\it Linearity.}\, For arbitrary operators $A$
and $B$\,,
$(A+B)|\Gamma_r\rangle=A|\Gamma_r\rangle+B|\Gamma_r\rangle$\,.
 \end{enumerate}

\subsection{Propagating operator and irreducible evolution operator}
\label{UE}

From now on we consider closed systems namely $F_{\rm ext}^\mu
=0$\,. In the absence of interactions, Eq.~(\ref{Liouville1}) is
reduced into
\begin{eqnarray}
\left (\frac{\partial}{\partial \tau}- {\hat {\mathfrak L}}^0\right)
{\cal D}= 0  \,.
\label{freemotion}
\end{eqnarray}
In this case, the evolution is determined by the propagating
operator $\U^0(\tau)$\,, or equivalently, its resolvent $\R^0(z)$
which are defined as
\begin{equation}
\U^0(\tau) = e^{\tau {\hat {\mathfrak L}}^0}\,, \qquad
\R^0(z)=\frac{1}{-iz-{\hat {\mathfrak L}}^0} \,.
 \label{UR}
\end{equation}
They are related through
\begin{equation}
\U^0(\tau) = (2\pi)^{-1}\int_C dz \, e^{-iz\tau} \R^0(z)\,,
 \label{URrelation}
\end{equation}
where the contour $C$ lies above all the singularities of the
Laplace transform of ${\cal D}
$\,.

We further introduce the irreducible evolution operator ${\hat
{\mathfrak{E}}}(\tau)$ and its Laplace transform $\E(z)$\,:
\begin{eqnarray}
\E(z) &=& \sum_{n=0}^\infty \lambda^{n+1} \Li'\{\C\R^0(z)\Li'\}^n
\,, \label{irreducibleoperator}\\
{\hat {\mathfrak{E}}}(\tau) &=& (2\pi)^{-1}\int_C dz \, e^{-iz\tau}
\E(z) \,. \nonumber
\end{eqnarray}
Then, the axiom dynamical system that we will consider throughout
this work is defined as
such systems that satisfy the following\\
\\

\noindent Assumption 3.1. {\it The operators: $\V \E(z)\V $\,, $\V
\E(z) \R^0(z)\C $\,, $\C \R^0(z) \E(z)\V $ and $\C \R^0(z) + \C
\R^0(z)\E(z)\R^0(z)\C $ are regular and nonvanishing at $z=0$\,.} \\

\noindent We stress that this assumption merely introduces
restrictions on microscopic interactions. A heuristic example where
this assumption is applicable is a system composed of identical
particles which interact with each other through short-ranged
interactions.

\subsection{Evolution of the kinetic component of distribution vector}
\label{kinetic}

Under Assumption 3.1. the following two theorems are straightforward
generalizations of their Newtonian counterparts, and we
shall not present the proof here \cite{Balescu}.\\

Theorem 3.2. {\it The distribution vector $\overrightarrow{{\frak
D}}$ may be decomposed into the kinetic component ${\hat \Pi}_{\rm
k.} \overrightarrow{{\frak D}}$ and the nonkinetic component ${\hat
\Pi}_{\rm n.k.} \overrightarrow{{\frak D}}$\,, i.e.,
\begin{eqnarray}
\overrightarrow{{\frak D}} = {\hat \Pi}_{\rm k.}
\overrightarrow{{\frak D}} + {\hat \Pi}_{\rm n.k.}
\overrightarrow{{\frak D}}\,.
\label{decomposition}
\end{eqnarray}
The vacuum part of the former satisfies a closed hierarchy:
\begin{equation}
\left(\frac{\partial}{\partial \tau}-\V\G\V\right) \V{\hat \Pi}_{\rm
k.} \overrightarrow{{\frak D}}=0 \,, \label{generalkinetic}
\end{equation}
where the operator $\V\G\V$
is given by the following functional equation:
\begin{eqnarray}
\V\G\V &=& \V\Li\V + \int_0^\infty ds\, \V\GG(s)\V \exp(-s\V\G\V) \,, \nonumber\\
\V\GG(s)\V &=& (2\pi)^{-1}\int_C dz\, e^{-izs} \V\E(z)\R^0(z)\C
\lambda \Li' \V \,.\nonumber\\
\label{VGV}
\end{eqnarray}
}\\

\noindent {\it Remark.} The operators ${\hat \Pi}_{\rm k.}$ and
${\hat \Pi}_{\rm n.k.}$ are $\tau$-independent nonlinear functionals
of the interacting Liouvillian $\lambda \Li'$\,. However, their
explicit expressions are not needed for subsequent analysis and,
therefore, we shall not present them here. Then, carrying out the
Laplace transform [denoting the Laplace transform of $\GG(s)$ as
$\GG_z$] and substituting Eq.~(\ref{irreducibleoperator}) into
Eq.~(\ref{VGV}), we find
\begin{equation}
\V\GG_z\V= \sum_{n=1}^\infty \lambda^{n+1}\V \Li' [\C\,
\R^0(z)\Li']^n\V \,. \label{G}
\end{equation}
Notice that the operator $\V\G\V$ depends on the given particle
world lines through the interacting Liouvillian $\lambda \Li'$\,.

For the correlation part, as shown in the following theorem, it is
fully determined by the vacuum part:\\

\noindent {Theorem 3.3.} {\it The kinetic component of the
distribution vector ${\hat \Pi}_{\rm k.} \overrightarrow{\mathfrak{D}}$
satisfies
\begin{eqnarray}
\C {\hat \Pi}_{\rm k.} \overrightarrow{\mathfrak{D}} =
\int_0^{\infty} ds\!\!\int_0^s ds'\C {\hat {\mathfrak{U}}}^0 (s-s'){\hat {\mathfrak{E}}}(s') \nonumber\\
\times \exp(-s\V\G\V) {\hat \Pi}_{\rm k.}
\overrightarrow{\mathfrak{D}} \,. \qquad
    \label{correlation}
\end{eqnarray}}\\

Eqs.~(\ref{generalkinetic}) and (\ref{correlation}) constitute the
main equations of the manifestly covariant classical correlation
dynamics. Crucially, they differ from traditional correlation
dynamics \cite{Prigogine65,Balescu64,Balescu} in that both the
operators $\G\,, {\hat {\mathfrak{E}}}\,, {\hat \Pi}_{\rm k.}$ and
the distribution vector $\overrightarrow{\mathfrak{D}}$ are
functionals of particle world lines. Such significant difference
causes additional conceptual and technical complications.

\section{Bogoliubov functional assumption}
\label{Bogoliubovhypothesis}

In the thermodynamic limit $N\rightarrow +\infty$ the kinetic
component obeys an infinite hierarchy. In this section we show that
such an infinite hierarchy may be reduced into a closed highly
nonlinear equation which, remarkably, admits a solution uniquely
determining the physical distribution function of full many-body
systems. In doing so we achieve a manifestly covariant version of
the Bogoliubov functional assumption.

\subsection{Hakim-Israel-Kandrup trial solution}
\label{stationary}

For the convenience below we introduce the mapping ${\cal X}$ which
uniquely converts a representation point in $\mu$ space into a world
line. It is defined as follows:
\begin{eqnarray}
{\cal X} :\, (x,p) \in \mu \mapsto {\cal X}_{(x,p)} \equiv x[s]
\label{T}
\end{eqnarray}
with the world line $x[s]$ satisfying\\
\\
\noindent (i)\, $ x|_{s=0} = x,\,
\frac{dx^\mu}{ds}\big|_{s=0} = p^\mu /m $\,;\\
\noindent (ii)\, $\forall \,\, s,\,\, \sqrt{\frac{dx^\mu}{ds}{\frac{dx_\mu}{ds}}} \leq 1\,$\,;\\
\noindent (iii) $d^4xd^4p$ is invariant along the phase trajectory: $(x[s],m\frac{d}{ds}x[s])$\,. \\
\\
\noindent The property (i) implies that $x[s]$ is a $1$-dimensional
submanifold of ${\cal M}^4$ which passes through $x$ with the
$4$-momentum vector $p$\,, (ii) implies that the
world line ${\cal X}_{(x,p)}$ preserves the causality, and (iii) implies that
particles move in the way as exposed to fictitious external field
which introduces ``conservative forces''.
Notice that the details of ${\cal X}$\,, although complicated generally, are
unimportant at this stage.

Eq.~(\ref{generalkinetic}) is a closed hierarchy of ${\hat \Pi}_{\rm
k.} \overrightarrow{{\frak D}}$\,. Thus, the latter as a whole may
be regarded as a new distribution vector, denoted as
$\overrightarrow{{\frak D}}^0 \equiv {\hat \Pi}_{\rm k.}
\overrightarrow{{\frak D}}$\,, which possesses the similar structure
as $\overrightarrow{{\frak D}}$\,, i.e., $\overrightarrow{{\frak
D}}^0\equiv (\{{\cal D}^0_1\} \,, \{{\cal D}^0_2\} \,, \cdots) $\,.
In this part we come to study a particular solution of
$\overrightarrow{{\frak D}}^0$ with the general component as
\begin{eqnarray}
&& {\cal D}^0_s(i_1, \cdots
\,,i_s;x_{i_1}[\varsigma],\cdots,x_{i_s}[\varsigma]) \nonumber\\
&=& {\cal
D}^0_s(i_1, \cdots \,,i_s;{\cal X}_{(x_{i_1},p_{i_1})},\cdots,{\cal X}_{(x_{i_s},p_{i_s})})
\nonumber\\
&\equiv & {\cal
D}^0_s(i_1, \cdots \,,i_s;{\cal X}_{i_1},\cdots,{\cal X}_{i_s})\,.
\label{component}
\end{eqnarray}
Also, we define the distribution vector $\overrightarrow{{\frak
D}}_\infty$ as follows:
\begin{eqnarray}
N^j \langle \Gamma_j(i_1,\cdots \,,i_j)|\overrightarrow{{\frak
D}}^0\equiv \langle \Gamma_j(i_1,\cdots \,, i_j)|\overrightarrow{{\frak
D}}_\infty \,, \nonumber\\
\qquad \forall \,\,  \langle \Gamma_j(i_1,\cdots \,,
i_j)|\,, j\geq 1\,. \qquad \qquad
\label{thermodynamiclimit}
\end{eqnarray}

In Ref.~\cite{Hakim67}--in an implicit manner--Hakim noticed that
in order to derive kinetic equations the distribution functions
in the $8$-dimensional $\mu$ phase space have to functionally depend on the
particle world line, and the latter merely relies on the phase coordinates in
the single particle $\mu$ phase space. In the
work of Israel and Kandrup \cite{Israel84,Kandrup84} this concept was reinforced
and explicitly formulated. Such an idea lies at the heart of Eq.~(\ref{component}). For this
reason we may call $\overrightarrow{{\frak D}}_0$ with the components given
by Eq.~(\ref{component}) the
Hakim-Israel-Kandrup (HIK) trial solution. From the HIK trial
solution an important fact follows: In general, to determine
expression of Eq.~(\ref{generalkinetic}) explicitly requires
calculations of the general matrix element $\langle
i_1|\cdots|i_{j}| \exp\{\tau \V\G\V\} |i_1|\cdots|i_{j'}\rangle$
with $j\leq j'$\,, where the sequence $i_1\cdots i_{j\, {\rm or}\,
j'}$ satisfies $i_s\neq i_{s'}\,, \forall\, s\neq s'$\,. On one
hand, $\exp\{\tau \V\G\V\}$ explicitly depends on the particle world
lines given by the acted distribution vector.
On the other hand, the particles, though having identical phase coordinates, may
still be ``distinguished'' through
their world lines which generally are coupled to each other.
Consequently, the quantity: $\langle i_1|\cdots|i_{j}| \exp\{\tau
\V\G\V\} |i_1|\cdots|i_{j'}\rangle \langle i_1|\cdots|i_{j'}|
\overrightarrow{{\frak D}}^0$ is sensitive to particles
$i_{j+1}\cdots i_j$ in the intermediate states. The substantial
simplification introduced by the HIK trial solution is just to
remove this sensitivity by decoupling the particle world lines. That is,
\begin{eqnarray}
&& \langle i_1|\cdots|i_{j}| \exp\{\tau \V\G\V\}
|i_1|\cdots|i_j|i_{j+1}|\cdots |i_{j'}\rangle \nonumber\\
&& \qquad \qquad \qquad \times \langle
i_1|\cdots|i_j|i_{j+1}|\cdots |i_{j'}| \overrightarrow{{\frak
D}}^0\nonumber\\
& = & \langle i_1|\cdots|i_{j}| \exp\{\tau \V\G\V\}
|i_1|\cdots|i_j|i'_{j+1}|\cdots |i'_{j'}\rangle \nonumber\\
&& \qquad \qquad \qquad \times \langle
i_1|\cdots|i_j|i'_{j+1}|\cdots |i'_{j'}| \overrightarrow{{\frak
D}}^0\,, \label{particlesymmetry}
\end{eqnarray}
where the particle groups $(i_{j+1},\cdots \,,i_{j'})$ and
$(i'_{j+1},\cdots \,, i'_{j'})$ are not identical.

For the HIK trial solution the factorization theorem below shows
that Eq.~(\ref{generalkinetic}) may be reduced into a single closed
equation of
reduced one-body distribution function, which is the generalization
of Clavin's theorem in the Newtonian physics
\cite{Clavin72}. \\
\\
\noindent Proposition 3.4. {\it In the limit $N\rightarrow +\infty$
the infinite hierarchy (\ref{generalkinetic}) may be reduced into a
single closed equation of reduced one-body distribution function as
follows:
\begin{eqnarray}
&& \left\{ \frac{\partial}{\partial \tau} + u^\mu_1
\partial_{\mu 1} - \int d2 \,\lambda \LI'_{12}\,
{\tilde {\cal D}}(2;{\cal X}_2)\right\} {\tilde {\cal D}}(1;{\cal X}_1) \nonumber\\
&=& \sum_{j \geq 2}\int \! d2 \! \cdots \! \int\! dj \langle 1 |\V (\G -
\Li) \V  |1|\cdots |j\rangle \prod_{s=1}^j {\tilde {\cal
D}}(s;{\cal X}_s).\nonumber\\
\label{generalkinetic5}
\end{eqnarray}
Furthermore, the stationary solution with respect to the
$\tau$-parametrized evolution above may be reduced into the
following closed equation:
\begin{eqnarray}
\left\{u^\mu_1 \partial_{\mu 1}  - \int_{\Sigma_2 \otimes U^4_2 }
d\Sigma_{\mu 2} d^4 p_2u^\mu_2\lambda \LI'_{12}f(2)
\right\} f(1) = \mathbb{K} [f] \nonumber\\
\label{reducedonebody}
\end{eqnarray}
with $f\in {\cal H}$ a physical
one-body distribution and $\mathbb{K}$ a nonlinear functional of $f$\,.}\\

\begin{widetext}
{\it Proof.} Let us consider the HIK trial solution. We may formally
solve the hierarchy (\ref{generalkinetic}) and project the solution
$\overrightarrow{{\frak D}}^0 (\tau)$ to the one-body vacuum state
say $\langle i |$\,. Consequently, we obtain (In order to make the
formula compact for the moment we suppress all the arguments of
$\overrightarrow{{\frak D}}^0$ except the proper time $\tau$
parametrizing the global evolution.)
\begin{eqnarray}
\langle i | \overrightarrow{{\frak D}}^0 (\tau)
= \langle i | \exp\{\tau \V\G\V\} |i\rangle\langle i|
\overrightarrow{{\frak D}}^0 (0)
+\sum_{j=1}^N \sum_{i_1\cdots i_j=1}^N\, \langle i | \exp\{\tau
\V\G\V\} |i|i_1|\cdots|i_j \rangle\langle
i|i_1|\cdots|i_j|\overrightarrow{{\frak D}}^0(0)\,,
\label{generalkinetic3}
\end{eqnarray}
where the sequence $i i_1\cdots i_j$ satisfies $i_s\neq i\,,
\forall\, s$ and $i_s\neq i_{s'}\,, \forall\, s\neq s'$\,. Without
loss of generality the sequence is ordered in the way that it starts
from $i$\,. Because of the particle symmetry namely
Eq.~(\ref{particlesymmetry}) we may simplify
Eq.~(\ref{generalkinetic3}) as (Without loss of generality we set
$i=1$\,.)
\begin{eqnarray}
\langle 1 |\overrightarrow{{\frak D}}^0 (\tau)  = \sum_{j = 1}^N\,
\frac{(N-1)!}{(N-j)!}\, \langle 1 |\exp\{\tau
\V\G\V\}|1|2|\cdots|j\rangle\langle
1|2|\cdots|j|\overrightarrow{{\frak D}}^0(0)\,,
\label{averagekinetic1}
\end{eqnarray}
where the combinatorial factors arises from the particle
symmetry. Notice that the matrix element of the operator $\exp\{\tau
\V\G\V\}$ is well defined in the limit $N\rightarrow +\infty$
because the world lines are given by the mapping ${\cal X}$\,.
Multiplying both sides by $N$ and taking
Eq.~(\ref{thermodynamiclimit}) into account we find
\begin{eqnarray}
\langle 1|\overrightarrow{{\frak D}}_\infty (\tau) = \sum_{j =
1}^\infty\,
 \langle 1 |\exp\{\tau \V\G\V\} |1|2|\cdots|j\rangle \,
\langle 1|2|\cdots|j|\overrightarrow{{\frak D}}_\infty(0)
\label{averagekinetic2}
\end{eqnarray}
from Eq.~(\ref{averagekinetic1}).

In general, for finite $N$ projecting $ \overrightarrow{{\frak D}}^0
(\tau)$ to a vacuum state, say $\langle i_1|\cdots|i_j|\,, 1<j<N$ we
obtain
\begin{eqnarray}
\langle i_1|\cdots|i_j |\overrightarrow{{\frak D}}^0 (\tau) &=&
\langle i_1|\cdots|i_j | \exp\{\tau \V\G\V\} \V
\overrightarrow{{\frak D}}^0(0) \nonumber\\
&=& \sum_{n=j}^{N} \sum_{i_{j+1}\cdots i_n=1}^N\, \langle
i_1|\cdots|i_j | \exp\{\tau \V\G\V\} |i_1|\cdots
|i_j|i_{j+1}|\cdots|i_n\rangle\langle
i_1|\cdots|i_j|i_{j+1}|\cdots|i_n|\overrightarrow{{\frak D}}^0(0)\,,
\label{averagekinetic11}
\end{eqnarray}
where for fixed $n$ the sequence $i_1\cdots i_n$ satisfies $i_s\neq
i_{s'}\,, \forall\, s\neq s'\,, 1\leq s, s'\leq n$\,. In the second
equality for the intermediate vacuum state the particle order again
needs to be distinguished and, without loss of generality, we set
the leading $j$ particles to be $i_1i_2\cdots i_j$\,. Because of the
particle symmetry with the limit $N\rightarrow +\infty$ taken we
obtain (setting $i_k=k\,, k=1,\cdots\,, j$)
\begin{eqnarray}
&& \langle 1|\cdots |j|\overrightarrow{{\frak D}}_\infty (\tau) \nonumber\\
&=&\sum_{n=j}^\infty \langle 1|\cdots|j | \exp\{\tau \V\G\V\}
|1|\cdots|n\rangle \langle 1|\cdots|n| \overrightarrow{{\frak
D}}_\infty (0)
\equiv
R
\label{averagekinetic3}
\end{eqnarray}
or equivalently,
\begin{eqnarray}
\overrightarrow{{\frak D}}_\infty (\tau) = \exp\{\tau \V\G\V\}
 \overrightarrow{{\frak D}}_\infty (0)
\label{averagekinetic4}
\end{eqnarray}
by noticing $\langle 1|\cdots|j | \exp\{\tau \V\G\V\}
|1|\cdots|n\rangle =0 $ for $j> n$\,.

We then come to analyze $R$\,. For this purpose we consider the
diagrammatical presentation of general matrix element: $\langle
1|\cdots|j | \exp\{\tau \V\G\V\} |1|\cdots|n\rangle \,, j\leq n$
describing the transition from the intermediate to final vacuum
state. Fully parallel to the Newtonian physics \cite{Balescu},
starting from Eq.~(\ref{VGV}) and employing the correlation pattern
representation of operators $\U^0$ and $\lambda \Li'$ (see
Appendix~\ref{partitionLiouville}) one may show that diagrams
representing the transition element are composed of $j$ disconnected
clusters. Each of them involves a particle group ${\rm P}_i$
($i=1,2,\cdots\,, j$) with $1+j_i$ particles ($j_i \geq 0$) which
are labeled as $i$ and $s_i,s_i+1,\cdots,s_i+j_i\,,
j+1=s_1<\cdots<s_j\leq n$\,. ${\rm P}_i$'s constitute a partition of
$(1,\cdots\,, n)$\,:
\begin{eqnarray}
&& {\rm P}_1\cup\cdots \cup {\rm P}_j=(1,\cdots,n)\,, \qquad i\in
{\rm
P}_i \,, \nonumber\\
&& \qquad {\rm P}_i\cap {\rm P}_{i'}=\emptyset, \qquad  \forall\,
i\neq i'\,.\label{averagekinetic5}
\end{eqnarray}
These clusters share a common diagrammatical feature: They start
from $1+j_i$ disconnected particle lines at the right-most side,
which gives the vacuum state $|0_i({\rm P}_i) \rangle\equiv
|i|s_i|s_i+1|\cdots|s_i+j_i\rangle$\,. Upon propagating to the left
particles $s_i,s_i+1,\cdots\,,s_i+j_i$ are annihilated and
eventually at the left-most side only particle $i$ is left giving
the vacuum state $\langle i|$\,. Thus, we factorize the matrix
element $R$ into $j$ components:
\begin{eqnarray}
R &=& \sum_{n=j}^\infty \sum_{{\rm P}_1\cdots {\rm P}_j}
\prod_{i=1}^j\, \langle i|\exp\{\tau \V\G\V\} |0_i({\rm P}_i)
\rangle \langle 0_i({\rm P}_i) |
\overrightarrow{{\frak D}}_\infty(0)\nonumber\\
&=& \sum_{s_1=1}^\infty\cdots \sum_{s_j=1}^\infty\prod_{i=1}^j\,
\langle i|\exp\{\tau \V\G\V\} |i_1|\cdots|i_{s_i}\rangle \langle
i_1|\cdots|i_{s_i} | \overrightarrow{{\frak D}}_\infty
(0)\nonumber\\
&=& \prod_{i=1}^j\sum_{s_i=1}^\infty\, \langle i|\exp\{\tau \V\G\V\}
|i_1|\cdots|i_{s_i}\rangle \langle i_1|\cdots|i_{s_i} |
\overrightarrow{{\frak D}}_\infty (0)\,,
 \label{averagekinetic6}
\end{eqnarray}
where in the second equality we use the particle symmetry
namely Eq.~(\ref{particlesymmetry}) to make the change of variables: ${\rm
P}_i \rightarrow i_1, i_2,\cdots\,,i_{s_i} $ in such a way that $i_1
\equiv i$ and $i_s \neq i_{s'}\,, s\neq s'$\,.

With Eq.~(\ref{averagekinetic6}) substituted
Eq.~(\ref{averagekinetic3}), together with
Eq.~(\ref{averagekinetic2}) then gives
\begin{eqnarray}
\langle 1|\cdots |j|
\overrightarrow{{\frak D}}_\infty (\tau)
=
\prod_{i=1}^j\sum_{s_i=1}^\infty\, \langle i|\exp\{\tau \V\G\V\}
|i_1|\cdots|i_{s_i}\rangle \langle i_1|\cdots|i_{s_i} |
\overrightarrow{{\frak D}}_\infty (0)\,, \qquad \forall\, j\geq 1\,.
\label{averagekinetic7}
\end{eqnarray}
\end{widetext}
Such an infinite hierarchy is solved by
\begin{eqnarray}
&& \langle 1|\cdots |j| \overrightarrow{{\frak D}}_\infty (\tau) =
\prod_{s=1}^j\, N {\cal D}^0_1(s;{\cal X}_s)
\nonumber\\
& \equiv & \prod_{s=1}^j\, {\tilde {\cal
D}}(x_s,p_s,\tau_s+\tau;{\cal X}_{(x_s,p_s)})\,, \qquad \forall \,
j\geq 1 \,, \label{factorization}
\end{eqnarray}
where in the last equality we have retrieved all the arguments of
the distribution functions. Indeed, with such a solution inserted
the infinite hierarchy (\ref{averagekinetic7}) is reduced into a
single closed equation as follows:
\begin{eqnarray}
&& {\tilde {\cal D}}(1;{\cal X}_1)
\label{averagekinetic8}\\
&=& \sum_{j=1}^\infty\, \langle 1| \exp\{\tau \V\G\V\} |1|\cdots|j
\rangle\, \prod_{s=1}^j {\tilde {\cal D}}(s;{\cal X}_s)|_{\tau=0}
\,. \nonumber
\end{eqnarray}
Taking the derivative with respect to $\tau$ we obtain
\begin{eqnarray}
&& \frac{\partial}{\partial \tau}{\tilde {\cal D}}(1;{\cal X}_1)
= \sum_{j=1}^\infty\, \langle 1| \V\G\V \exp\{\tau \V\G\V\}
\overrightarrow{{\frak D}}_\infty (0)  \nonumber\\
&=& \sum_{j=1}^\infty\, \langle 1| \V\G\V |1|\cdots |j\rangle
\langle 1|\cdots |j| \exp\{\tau \V\G\V\} \overrightarrow{{\frak
D}}_\infty (0)  \nonumber\\
&=& \sum_{j=1}^\infty\, \langle 1| \V\G\V |1|\cdots |j\rangle
\prod_{s=1}^j\, {\tilde {\cal D}}(s;{\cal X}_s)\,,
\label{averagekinetic12}
\end{eqnarray}
which is the differential form of Eq.~(\ref{averagekinetic8}). We
thus prove the first part of the proposition.

Let us now fix a spacelike $3$-surface $\Sigma_i$ for particle $i$
and denote the coordinate as $\sigma_i$\,. Then, the stationary
${\tilde {\cal D}}(i;{\cal X}_i)$--with respect to the
$\tau$-parametrized evolution--assumes the following form:
\begin{equation}
{\tilde {\cal D}}(i;{\cal X}_i) = {\tilde f}(\sigma_i,p_i;{\cal X}_{(x_i,p_i)}
(\tau_i
-\varsigma_i))\,.
    \label{stationarysolution}
\end{equation}
Here $\varsigma_i$ is the proper time when the world line ${\cal
X}_{(x_i,p_i)}$ passes through the phase point $(x_i,p_i)$\,, and
${\tilde f}(\sigma_i,p_i;{\cal X}_{(x_i,p_i)}(s))$ is a distribution
function peaking at the spacelike $3$-surface $\Sigma_i$\,.
Inserting Eq.~(\ref{stationarysolution}) into
Eq.~(\ref{generalkinetic5}) and integrating out $\varsigma_i\,,
i\geq 2$ and $\tau_1$\,, we obtain a general kinetic equation
(\ref{reducedonebody}) with the collision integral given by
\begin{eqnarray}
\mathbb{K}[f] = \sum_{j \geq 2} \, \int_{\Sigma_2 \otimes U^4_2 }
d\Sigma_{\mu 2} d^4 p_2 u^\mu_2\cdots \int_{\Sigma_j \otimes U^4_j
} d\Sigma_{\mu j} d^4 p_j u^\mu_j \nonumber\\
\times \langle 1 |\V (\G - \Li) \V
|1|\cdots |j\rangle \, \prod_{i=1}^j\, f(i) \,,\qquad \qquad
\label{averagekinetic10}
\end{eqnarray}
where
\begin{equation}
f(x_i,p_i) = \int ds\, {\tilde f}(\sigma_i,p_i;{\cal X}_{(x_i,p_i)}
(s)) \in {\cal H}_i \,.
    \label{physicaldistribution}
\end{equation}
Notice that in Eq.~(\ref{averagekinetic10}) the two-body interacting Liouvillian
$\lambda\LI'_{ij}(x_i,p_i;x_j,p_j)$ is a functional of the world
lines ${\cal X}_{(x_i,p_i)}$ and ${\cal X}_{(x_j,p_j)}$\,. The second
part
of the proposition then follows. Q.E.D.\\

{\it Remark.} (i) By definition the distribution function
(\ref{stationarysolution}) satisfies ${\tilde {\cal D}}|_{\tau_i
\rightarrow \pm \infty}=0$ and thereby is compatible with Lemma 2.2.
$\tilde f$ may take a particular form: $
f_{\rm K}(x_i,p_i)\, \delta (\tau_i-\varsigma_i)$ with $f_{\rm
K}(x_i,p_i)\in {\cal H}_i$\,, which was given--in an implicit
manner--in Ref.~\cite{Kandrup84}. (ii) An equation similar to
Eq.~(\ref{generalkinetic5}) has been obtained by Hakim
\cite{Hakim67} for dilute gases with scalar and electromagnetic
interactions. There, the equation is derived under the weak coupling
approximation (namely the second order interaction expansion). In
contrast, provided that the force is conservative
Eq.~(\ref{generalkinetic5}) presented here is exact including all
the higher order (three-, four-body, etc.) correlations. It
encompasses a route to go beyond the weak coupling approximation for
dilute gases, and is essential to the justification of the
relativistic Boltzmann equation (to be detailed in Paper II).

\subsection{Physical correlation functions}
\label{physicalcorrelation}

The solution of Eq.~(\ref{reducedonebody}) fully determines the
entire physical correlation functions. Indeed, let us replace ${\hat \Pi}_{\rm k.}\overrightarrow{\mathfrak{D}}$
in Eq.~(\ref{correlation}) with $\overrightarrow{{\frak D}}^0$\,,
the components of which are given by Eq.~(\ref{component}).
Taking into account Eq.~(\ref{stationarysolution}), by straightforward calculations
we may find the hierarchy of
physical correlations in the limit: $N\rightarrow +\infty$\,,
denoted as $\C \overrightarrow{\mathfrak{N}}_\infty$\,, to be
\begin{eqnarray}
\C \overrightarrow{\mathfrak{N}}_\infty &=&
\int_0^{\infty}ds\!\!\int_0^s ds'\C {\hat {\mathfrak{U}}}^0 (s-s'){\hat {\mathfrak{E}}}(s') \V \nonumber\\
&& \times \exp(-s\V\G\V) \overrightarrow{\mathfrak{N}}_\infty \,.
    \label{correlation1}
\end{eqnarray}
In the derivation we notice that the vacuum part of
the physical distribution vector
is given by
\begin{equation}
\langle i_1|\cdots |i_n | \overrightarrow{\mathfrak{N}}_\infty = \prod_{j=1}^n f(i_j)
    \label{correlation2}
\end{equation}
for arbitrary vacuum state $\langle
i_1|\cdots |i_n |$\,.

Alternatively, we may project Eq.~(\ref{correlation1}) to the
general $j$-particle correlation pattern $\Gamma_j (i_1,\cdots\,,
i_j)\neq 0_j(i_1,\cdots\,,i_j)$ and arrive at
\begin{eqnarray}
\langle \Gamma_j
|\overrightarrow{\mathfrak{N}}_\infty
= \sum_{n=j}^\infty \int_0^{\infty} ds\!\!\int_0^s ds'\, \prod_{j=1}^n f(i_j) \qquad \qquad \qquad
\label{correlationresult2}\\
\times \langle\!\langle \Gamma_j
| \C {\hat {\mathfrak{U}}}^0 (s-s'){\hat {\mathfrak{E}}}(s') \V
\exp(-s\V\G\V)|i_1|\cdots |i_n\rangle\!\rangle \,, \nonumber
\end{eqnarray}
where $\langle\cdot\rangle$ stands for the replacement:
$\int di \rightarrow \int_{\Sigma_i \otimes U^4_i } d\Sigma_{\mu i}
d^4 p_i\, u^\mu_i$ for the annihilation vertex [see
Eq.~(\ref{BBGKY5})].

This exact formula provides a principle to calculate arbitrary
physical correlation functions, provided that the physical one-body
distribution function namely $f(x,p)$ is given.
Eqs.~(\ref{reducedonebody}) and (\ref{correlationresult2}) show that
(after transient processes) physical correlation functions relax
into functionals of physical one-body distribution function. The
latter obeys closed kinetic equation (\ref{reducedonebody}) from
which a relativistic hydrodynamic description stems. Thus, we
justify the manifestly covariant the Bogoliubov functional
assumption.

\section{Some remarks on further applications}
\label{remark}

It has been a long standing problem to reconcile statistical
mechanics and relativity principles. In recent years this subject
has become fundamentally important to studies in many fields as
mentioned in the introductory section. In 1984 Israel and Kandrup
made substantial progresses in this direction
\cite{Israel84,Kandrup84}. There, the authors formulated manifestly
covariant classical nonequilibrium statistical mechanics and
successfully applied it to relativistic plasmas with electromagnetic
interactions. In particular, the relativistic Landau equation
\cite{Belyaev57} was justified at the full microscopic level. It
differed remarkably from earlier attempts
\cite{Prigogine65,Klimontovich60} in that the theory is manifestly
covariant at each step of the manipulation. As such, the formulated
nonequilibrium statistical mechanics (particularly various
approximations unavoided on top of it) is guaranteed to suit
relativity principles automatically. That impeding further
development lies, as pointed out in the original paper
\cite{Kandrup84}, in that the Israel-Kandrup formalism fails to go
beyond the weak coupling approximation and to capture the collective
effects. (It is important that the latter heals the well-known
logarithmic divergence of the Landau equation.) One of the main
purposes of this series of papers, indeed, is to attack this
difficult problem.

In this paper we start from the single-time Liouville equation
(\ref{Liouville1}) and derive rigorously the manifestly covariant
closed nonlinear equation (\ref{reducedonebody}), satisfied by the
reduced one-body distribution function. (The latter indicates that a
necessary condition to establish the irreversibility is the
introduction of the thermodynamic limit $N\rightarrow +\infty$\,.)
The solution to Eq.~(\ref{reducedonebody}), in turn, determines the
hierarchy of correlations namely Eq.~(\ref{correlationresult2}). We
remark that although Eq.~(\ref{reducedonebody}) is exact, it differs
from the usual relativistic kinetic equations
\cite{Belyaev57,Cercigani02} in the non-Markovian feature rendering
the failure of $H$-theorem \cite{Prigogine66}. This feature was
known long time ago in the Newtonian physics (for example, the
Prigogine-Resibois master equation \cite{Prigogine61}). It may, as
pointed out in a pioneering study \cite{Prigogine66}, play essential
roles in understanding the infrared divergence of the collision
integral of self-gravitating systems (which we will discuss below).
It is only beyond the correlation scale (provided that the two-body
correlation is short-ranged) that the general equation
(\ref{reducedonebody}) converges to a usual kinetic equation. (The
very latter leads to a macroscopic hydrodynamic description.) In
Paper II we exemplify this scenario in the case of relativistic
plasmas with electromagnetic interactions. There, it will become
clearer that Eqs.~(\ref{reducedonebody}) and
(\ref{correlationresult2}) allow one to overcome the difficulty
encountered in the Israel-Kandrup formalism. Furthermore, we manage
to justify the manifestly covariant Boltzmann equation
\cite{Cercigani02,vanLeeuwen80}, and are able to explore--at the
quantitative level--the origin of the logarithmic divergence of the
relativistic Landau equation. They both are extremely hard to be
achieved within the Israel-Kandrup formalism.

It should be stressed that the general theory, under Assumption 2.1.
and 3.1, is rigorous. These two assumptions essentially require that
(classical) chaotic dynamics sufficiently develops for the
underlying relativistic many-body system. Similar to their Newtonian
analogues to justify them mathematically is an extremely hard task
and is unnecessary at this stage. Rather, from the practical
viewpoint the theory presented here may be applicable to a large
class of systems where the interacting force preserves the
mass-shell constraint and, moreover, does not depend on the
acceleration of the acted particle. A heuristic example is a system
composed of identical particles which interact with each other
through a weak enough massive scalar field. In this case the
interacting force is
\begin{eqnarray}
&&  F^\mu_{ij}(x_i,p_i) = \lambda_0\,
(\eta^{\mu\nu}-u^\mu_i u^\nu_i) \partial_{\nu i} \Phi_{ij}(x)\big|_{x=x_i} \,, \\
&&  (\partial_\mu\partial^\mu + \kappa^2) \, \Phi_{ij}(x) = -4\pi
  \lambda_0 \int d\tau_j \delta^{(4)}(x-x_j(\tau_j)) \,,
\nonumber
\end{eqnarray}
where $|\lambda_0 \Phi_{ij}|\ll m$\,,  the interaction strength is
$\propto \lambda_0^2$\,, and $\kappa$ is the mass of the scalar
field. Notice that the interacting force preserves the mass-shell
constraint, i.e., $u_i \cdot F_{ij}(x_i,p_i)=0$\,, but does not
satisfy the conservative condition namely
Eq.~(\ref{forcecondition}). Nevertheless it turns out that the
general theory well applies in this case and, furthermore, the
entire program of Paper II may be readily carried over to this
system. In particular, because of the short-ranged nature of the
interacting force a well defined Landau collision integral results
under the weak coupling approximation.

\subsection{Relativistic self-gravitating systems}
\label{gravitation}

In this series of papers we limit ourselves to the special
relativity. The theoretic scope presented here may be further
extended to the general relativity and, thus, find important
applications in astrophysics. In particular, there are no
difficulties which impede generalizing the present manifestly
covariant classical correlation dynamics to self-gravitating systems
composed of relativistic star clusters or galaxies interacting
merely gravitationally. Since this subject is far beyond the scope
of this series of papers we here report briefly some preliminary
observations and leave the thorough analysis for future studies.

In formulating the classical correlation dynamics of relativistic
self-gravitating systems an additional difficulty arises, namely to
treat the ``gravitational force''. We here follow the prescription
of Israel and Kandrup \cite{Israel84}. First of all, we fix a
background geometry $({\cal M}^4, g_{\mu\nu}(x))$\,. The
$8$-dimensional $\mu$ phase space [compared with Eq.~(\ref{space})]
associated with particle, say $i$ is now defined as $\mu_i:
\{(x_i,p_i)|x_i \in {\cal M}^4\,, g_{\mu\nu}(x_i) p^{\mu}_i
p^{\nu}_i=m^2 \}$\,. The background metric $g_{\mu\nu}(x)$ is such
chosen that it solves some field equation with smoothed-out matter
distribution as the source. Crucially, given particle $i$ the
realistic path $x_i[\varsigma]$--dictated by underlying classical
many-body dynamics--is the geodesic generated by another metric
$g'_{\mu\nu}(x)$\,. It generally deviates from the geodesic in the
background geometry $({\cal M}^4, g_{\mu\nu}(x))$ (and, as such, do
not preserve the phase volume of $\mu_i$\,.) This very deviation is
driven by the difference of the Christoffel symbols
${\Gamma}^\lambda_{\mu \nu}$ and ${\Gamma'}^\lambda_{\mu \nu}$
(associated with the metrics $g_{\mu\nu}$ and $g'_{\mu\nu}$\,,
respectively), i.e., $\delta\Gamma^\lambda_{\mu \nu}(x)\equiv
{\Gamma'}^\lambda_{\mu \nu}(x)-\Gamma^\lambda_{\mu \nu}(x)$\,, and
describes the gravitational force completely. Provide that
$\delta\Gamma^\lambda_{\mu \nu}(x)$ varies over a scale much smaller
than the radius of the spacetime curvature, it may be determined by
some linear field equation.

To substantiate these observations we need to make the following
replacement in Eq.~(\ref{Liouville1})
($\Delta^i_{\mu\nu}=g_{\mu\nu}-u_{\mu i}u_{\nu i}$):
\begin{eqnarray}
{\hat {\mathfrak{L}}}^0 = -\sum_{i=1}^N \left[u_i^\mu
\partial_{\mu i} + m\Gamma^\lambda_{\mu \delta}(x_i)
u_{\lambda i}u_i^\delta \frac{\partial}{\partial p_{\mu i}}\right]\,,
\label{interactingLiouville2}\\
\lambda {\hat {\cal L}}'_{ij}=m \delta \Gamma^\lambda_{\nu
\delta}(i,j) \frac{\partial}{\partial p_{\mu i}}
\Delta^i_{\mu\lambda} u_i^\nu u_i^\delta
 + (i\leftrightarrow j)
\,,
\label{interactingLiouville1}
\end{eqnarray}
where in Eq.~(\ref{interactingLiouville1}) the second term is
obtained by exchanging the particle labels of the first term. Here
we notice that because of the linear field approximation mentioned
above the perturbed Christoffel symbol of particle $i$ namely
$\delta\Gamma^\lambda_{\mu \nu}(x_i)$ consists of $N-1$
contributions from all the other particles, and $\delta
\Gamma^\lambda_{\nu \delta}(i,j)$ is the one associated with
particle $j$\,.

Then, one may further proceed to formally carry out the entire
program presented here. In particular, we obtain, similar to
Eq.~(\ref{reducedonebody}), an exact closed nonlinear equation
satisfied by the one-body distribution function:
\begin{eqnarray}
\bigg\{u^\mu_1 \partial_{\mu 1} + m\Gamma^\lambda_{\mu \delta}(x_1)
u_{\lambda 1}u_1^\delta \frac{\partial}{\partial p_{\mu 1}}
\qquad \qquad \qquad \qquad \quad \nonumber\\
- \int_{\Sigma_2}\!d\Sigma_{\mu 2}\! \int\!\frac{d^4 p_2}{\sqrt {g}}
u_2^\mu \, \lambda \hL f(2)\bigg\}f(1) = \mathbb{K}_{\rm g}[f]\,,
\label{IK}
\end{eqnarray}
where
$f$ is normalized according to $\lim_{N\rightarrow +\infty}
N^{-1}\int_{\Sigma}\!d\Sigma_{\mu }\! \int d^4 p u^\mu f(x,p)=1$
with $\Sigma$ a spacelike $3$-surface, and $\mathbb{K}_{\rm g}[f]$
is the collision integral (The explicit form may be readily found
which is yet unnecessary for present discussions.). Because the
perturbed Christoffel symbol is weak enough to the leading order
$\lambda$-expansion Eq.~(\ref{IK}) then gives
\begin{eqnarray}
\left\{u^\mu_1 \partial_{\mu 1} + m\Gamma^\lambda_{\mu \delta}(x_1)
u_{\lambda 1}u_1^\delta \frac{\partial}{\partial p_{\mu 1}} \right
\}f(1) = 0\,,
\label{EinsteinBoltzmann1}
\end{eqnarray}
the solution to which self-consistently determines the background
geometry, i.e., $({\cal M}^4,g_{\mu \nu}(x))$ through the Einstein
field equation (${G^\mu}_\nu$ is a function of the Christoffel
symbol.):
\begin{eqnarray}
{G^\mu}_\nu =-8\pi \int\frac{d^4p}{\sqrt{g}} u^\mu p_\nu f(x,p) \,.
\label{EinsteinBoltzmann2}
\end{eqnarray}
Eqs.~(\ref{EinsteinBoltzmann1}) and (\ref{EinsteinBoltzmann2})
constitute the well-known Boltzmann-Einstein equations
\cite{Israel84}.

Let us further replace ${\cal X}_{(x,p)}$ [cf. Eq.~(\ref{T})] by the
phase trajectory generated by the geodesic in the background
geometry. Provided that the self-gravitating system is dilute, by
taking into account the so-called locality assumption we are able to
show that the collision integral is
\begin{eqnarray}
\mathbb{K}_{\rm g}[f] &=& \int \frac{d^4 p_2}{\sqrt {g}} \left(
\frac{\partial}{\partial p_1^\mu}
 -\frac{\partial}{\partial p_2^\mu}\right)
\epsilon^{\mu\nu}_{\rm
g}\left(\frac{\partial}{\partial
p_1^\nu}-\frac{\partial}{\partial p_2^\nu}\right) \nonumber\\
&& \qquad \qquad \qquad \times f(x_1,p_1)f(x_1,p_2) \,, \label{IKLandau}\\
\epsilon^{\mu\nu}_{\rm g} &=& 2m^4 [1-2(u_1\cdot u_2)^2]^2 \nonumber\\
&& \qquad \times \int \frac{d^4 k}{\sqrt {g}} \delta(k\cdot
u_1)\delta(k\cdot u_2) \frac{k^\mu k^\nu}{(k\cdot k)^2}
\nonumber
\end{eqnarray}
up to the $\lambda^2$ accuracy (namely the weak coupling
approximation to be detailed in Paper II). This is the
Israel-Kandrup collision integral \cite{Israel84} well-known to
astrophysicists.

The collision integral (\ref{IKLandau}) justifies that the classical
correlation dynamics can be well applied to relativistic
self-gravitating systems, which was questioned by Kandrup long time
ago \cite{Kandrup84}. It should be stressed, however, that such a
justification is at the level of formal manipulations. The serious
difficulty is that this collision integral suffers from the infrared
divergence, and may be amounted to the failure of the locality
assumption. (The ultraviolet divergence is due to the failure of the
weak coupling approximation which may be readily healed and, thus,
of no special interests \cite{Tian09}.) Such a divergence, like
plasmas \cite{Balescu}, finds its origin at the long-ranged nature
of gravitational forces, which was noticed by Chandrasekhar long
time ago \cite{Chandrasekhar}. However, in (Coulomb) plasmas the
Debye screening renders the effective potential short-ranged. As a
result, the two-body correlation is short-ranged justifying the
locality assumption, which results in a well defined kinetic
equation \cite{Balescu60}. (One of the central issue of Paper II,
indeed, is the detailed analysis of the analogue in the manifestly
covariant classical correlation dynamics.) This scenario
nevertheless breaks down in self-gravitating systems. Indeed,
experiences in Newtonian self-gravitating systems suggest that the
screening is dynamical \cite{Kukharenko94}. The general formalism
presented here may serve as a useful technique for exploring this
fundamental issue, which we leave for future studies.

\acknowledgements

I am deeply grateful to Q. K. Lu for numerous fruitful discussions
at the early stage of this work, and especially to S. L. Tian for
invaluable help. I also would like to thank M. Courbage and M. Garst
for useful conversations, and especially to C. Kiefer for his
interests and encouragements. This work is supported by Transregio
SFB 12 of the Deutsche Forschungsgemeinschaft and was partly done in
Institute of Henri Poincare.

\begin{appendix}

\section{BBGKY hierarchy of physical distribution functions}
\label{BBGKY2}

Let us introduce the concept of the physical distribution vector as
follows:
\begin{eqnarray}
\overrightarrow{{\frak N}} \equiv (\{{\cal N}_1\} \,, \{{\cal N}_2\}
\,, \cdots \,,\{{\cal N}_N \}\equiv {\cal N})\,. \label{Nvector}
\end{eqnarray}
The general component above is defined as
\begin{eqnarray}
&& {\cal N}_s(x_{i_1},p_{i_1}, \cdots
\,,x_{i_s},p_{i_s};x_{i_1}[\varsigma],\cdots,x_{i_s}[\varsigma])
\nonumber\\
&\equiv& \prod_{j=s+1}^N \,\int_{\Sigma_{i_j}\otimes U_{i_j}^4} d\Sigma_{\mu
{i_j}} d^4 p_{i_j} u^\mu_{i_j}\,\, {\cal N}
\label{Ncomponent}
\end{eqnarray}
for $1\leq i_1<\cdots <i_s \leq N\,, 1\leq s\leq N$\,. The
normalization of the component is given by
\begin{eqnarray}
\int_{\Sigma_{i_1}\otimes U_{i_1}^4} d\Sigma_{\mu
{i_1}} d^4 p_{i_1} u^\mu_{i_1}\cdots \int_{\Sigma_{i_s}\otimes U_{i_s}^4} d\Sigma_{\mu
{i_s}} d^4 p_{i_s} u^\mu_{i_s}\, \nonumber\\
\times {\cal N}_s(x_{i_1},p_{i_1},
\cdots
\,,x_{i_s},p_{i_s};x_{i_1}[\varsigma],\cdots,x_{i_s}[\varsigma])=1 \qquad
\label{normalization6}
\end{eqnarray}
for arbitrary spacelike $3$-surface $\Sigma_{i_j}$\,.

Similar to Sec.~\ref{reduceddistribution} from the Liouville
equation (\ref{Liouville2}) we obtain
\begin{equation}
{\hat {\mathfrak L}}\overrightarrow{{\frak N}}=0 \,.
\label{BBGKY4}
\end{equation}
Upon projecting it to the general $s$-particle state
$(i_1\,,\cdots\,, i_s|$ we use the same matrix elements except that
the annihilation vertex (see Appendix~\ref{partitionLiouville}) is
replaced by
\begin{eqnarray}
&& (i_1\,,\cdots\,, i_j| \lambda \Li' |i_1\,,\cdots\,, i_j,i_{j+1})
\label{BBGKY5}\\
&\rightarrow&  -\int_{\Sigma_{i_{j+1}}\otimes U_{i_{j+1}}^4}
d\Sigma_{\mu {i_{j+1}}} d^4 p_{i_{j+1}} u^\mu_{i_{j+1}}\,
\sum_{k=1}^j \, \lambda \LI'_{i_ki_{j+1}}\,.
\nonumber
\end{eqnarray}
Notice that the BBGKY hierarchy (\ref{BBGKY4}) is manifestly
covariant. If $\overrightarrow{{\frak N}}$ does not depend on
the particle world lines, then we recover a nonmanifestly covariant
BBGKY hierarchy by enforcing $t_1=\cdots =t_N
\equiv t$\,, which is perfectly legitimate by relativity principles.

\begin{figure}
\begin{center}
\leavevmode \epsfxsize=8cm \epsfbox{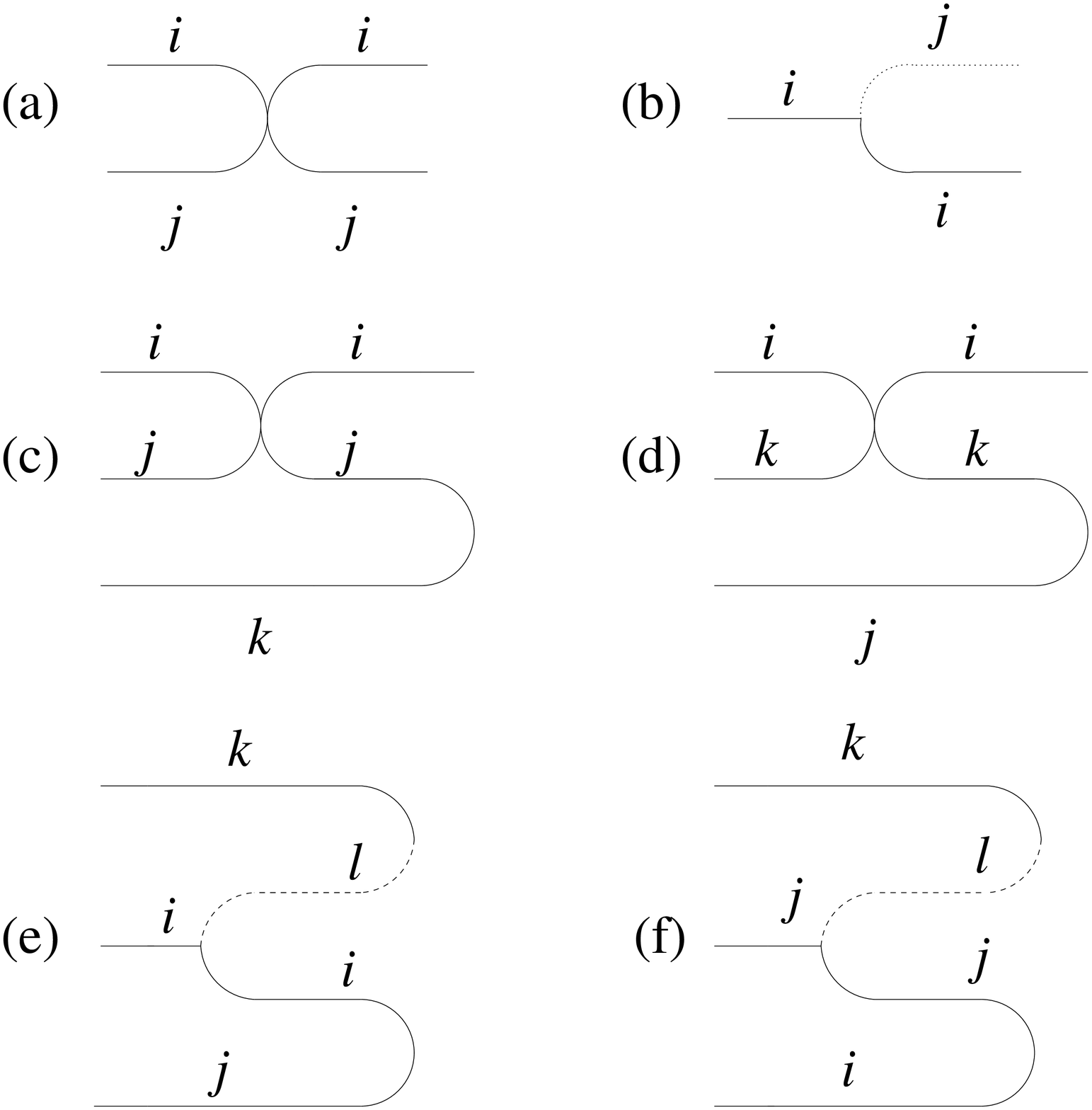}
\end{center}
\caption{(a) and (b): two types of interaction vertex, where dotted
line stands for the annihilated particle. (c)-(f): examples of
diagrams with $s'=3$\,.}
  \label{fig}
\end{figure}

\section{Liouvillian in correlation pattern representation}
\label{partitionLiouville}

In this appendix we will give the matrix elements of the free and
interacting Liouvillian in the correlation pattern representation.
For this purpose it is sufficient to know the matrix elements of the
operators $-u^\mu_i\partial_{\mu i}$ and $\lambda \LI'_{ij}$ because
of the linearity. Let us project the BBGKY hierarchy (\ref{BBGKY1}) to
the correlation pattern $\langle \Gamma_s(i_1,\cdots\,, i_s)|$ then
substitute the cluster expansion
(\ref{partition}) into it. Comparing with Eq.~(\ref{BBGKY1}) (setting $\lambda=0$) we find
that the operator $-u^\mu_i\partial_{\mu i}$ preserves a given
correlation pattern $|\Gamma_s(i_1,\cdots\,, i_s)\rangle$ (namely the
free motion). That is,
\begin{eqnarray}
\langle \Gamma_s(i_1,\cdots\,, i_s)| \, (-u^\mu_i\partial_{\mu i})\,
|\Gamma_s(i_1,\cdots\,,i_s)\rangle = -u^\mu_i
\partial_{\mu i}
\nonumber\\
\label{freematrixelement}
\end{eqnarray}
if $i \in (i_1,\cdots\,,i_s)$\,. All the other matrix elements vanish.

Similarly, we find that, given a two-particle correlation pattern
$|\Gamma_2(i,j)\rangle$\,, the two-body interacting Liouvillian
$\lambda \LI'_{ij}\,, i\neq j$ either preserves the particles
(Fig.~\ref{fig} (a)) or annihilates one particle via the manipulation of integration
(Fig.~\ref{fig} (b)). That is,
\begin{eqnarray}
\langle ij|\lambda \LI'_{ij}|\Gamma_2(i,j)\rangle &=& \lambda
\LI'_{ij}\,, \nonumber\\
\langle i|\lambda \LI'_{ij}|\Gamma_2(i,j)\rangle
&=&\int dj\, \lambda\LI'_{ij}\,, \nonumber\\
\langle j|\lambda
\LI'_{ij}|\Gamma_2(i,j)\rangle &=& \int di\, \lambda\LI'_{ij}\,,
\label{interactionmatrixelement}
\end{eqnarray}
and all the other matrix elements vanish. In general, using the
linearity we find that for matrix elements $\langle
\Gamma_{s'}(i_1,\cdots\,, i_{s'})|\lambda\Li'|\Gamma_{s}(j_1,\cdots\,,
j_{s})\rangle$ not to vanish it is necessary that $(j_1,\cdots\,, j_{s})
=(i_1,\cdots\,,
i_{s'})$ or $(i_1,\cdots\,, i_{s'},
i_{s'+1})$\,. Besides, provided that the correlation pattern
$|\Gamma_{s}(j_1,\cdots\,, j_{s})\rangle=|{\rm P}_1|{\rm P}_2|\cdots
|{\rm P}_j\rangle $ with
\begin{eqnarray}
{\rm P}_1 \cup\cdots \cup {\rm P}_j &=& (j_1,j_2,\cdots\,, j_s)\,, \nonumber\\
{\rm P}_i \cap {\rm P}_{i'}&=&\emptyset \,, \forall\, i\neq i' \,,
\label{partitioncorrelation}
\end{eqnarray}
then the matrix element is invariant under the particle permutation
if both particles are in the same subset. This statement
is also applicable to the correlation pattern
$\langle \Gamma_{s'}|$\,.

Table 1 presents the matrix elements with $s'=1,2$\,. For the matrix
elements $\langle \Gamma_{s'}(i_1,\cdots\,,
i_{s'})|\lambda\Li'|\Gamma_{s}(j_1,\cdots\,, j_{s})\rangle$ with
$s'>2$\,, we set up the diagrammatical rules as follows:

\begin{enumerate}
\item The correlation pattern $|\Gamma_{s}(j_1,\cdots\,,
j_{s})\rangle$ is given by the right-most part of the diagram. If
particles are correlated then we connect them.
\item For each particle draw a horizontal line which stands for
the propagation from the right to the left. If particle is
annihilated during the propagation it is drawn by a dotted line,
otherwise by a solid line.
\item Draw all diagrams with an interacting vertex which
may be either Fig.~\ref{fig} (a) or (b). The obtained diagram,
as a whole, gives a possible final correlation pattern which is
read out according to the following rule: If particles are connected
(irrespective of solid/dotted line) then they are correlated.
\item Compare each possible final correlation pattern with
$\langle \Gamma_{s'}(i_1,\cdots\,, i_{s'})|$\,. If they are not
identical then we assign the value zero to the corresponding
diagram.
\item Otherwise, depending on the type of the interaction vertex we assign the value
$\lambda\LI'_{ij}$ or $\int dk\,
\lambda\LI'_{ik}$ accordingly. Here $i,j,k$ are the particles
joining the vertex.
\item Summing up all the nonvanishing diagrams then
gives the matrix element.
  \end{enumerate}

\begin{widetext}
\begin{center}
Table 1. Matrix element $\langle \Gamma_{s'}|\lambda
\Li'|\Gamma_{s}\rangle$ with $s'=1,2$\,. All the
other matrix elements not listed here are zero.  \\
\vspace{0.3cm}
\begin{tabular*}{0.85\textwidth}{@{\extracolsep{\fill}} | c | c | c | }
  \hline
$\langle i| $& $\langle i|j|$ & $\langle ij|$  \\
  \hline
$\langle i |\lambda \Li'|i\rangle =0 $ &  $\langle i|j |\lambda \Li'|i|j\rangle =0 $ & $\langle ij |\lambda \Li'|i|j\rangle =\lambda\LI'_{ij} $ \\
$ \langle i |\lambda \Li'|i|j\rangle =\int dj \, \lambda\LI'_{ij} $
&
$ \langle i|j |\lambda \Li'|ij\rangle =0 $ & $ \langle ij |\lambda \Li'|ij\rangle =\lambda\LI'_{ij} $ \\
$ \langle i |\lambda \Li'|ij\rangle =\int dj \, \lambda\LI'_{ij} $ &
$ \langle i|j |\lambda \Li'|i|j|k\rangle =\int dk\, \lambda
(\LI'_{ik}+\LI'_{jk})  $ &
$ \langle ij |\lambda \Li'|i|j|k\rangle =0 $ \\
& $ \langle i|j |\lambda \Li'|i|jk\rangle =\int dk\,
\lambda\LI'_{jk}  $ &
$ \langle ij |\lambda \Li'|i|jk\rangle =\int dk\, \lambda\LI'_{ik} $ \\
& $ \langle i|j |\lambda \Li'|j|ik\rangle = \int dk\,
\lambda\LI'_{ik}  $ &
$ \langle ij |\lambda \Li'|j|ik\rangle =\int dk\, \lambda\LI'_{jk} $ \\
& $ \langle i|j |\lambda \Li'|k|ij\rangle =0  $ &
$ \langle ij |\lambda \Li'|k|ij\rangle =\int dk\, \lambda(\LI'_{ik}+\LI'_{jk}) $ \\
& $ \langle i|j |\lambda \Li'|ijk\rangle =0  $ & $ \langle ij
|\lambda \Li'|ijk\rangle =\int dk\, \lambda(\LI'_{ik}+\LI'_{jk}) $ \\
\hline
\end{tabular*}
\end{center}
\end{widetext}

For illustrations here we give some examples: $\langle ijk |
\lambda\Li'|i|jk\rangle = \lambda(\LI'_{ij}+\LI'_{ik})$
(Fig.~\ref{fig} (c) and (d)), $\langle ijk |
\lambda\Li'|ij|kl\rangle =\int dl\, \lambda (\LI'_{il}+\LI'_{jl})$
(Fig.~\ref{fig} (e) and (f)), and $\langle ijk |
\lambda\Li'|i|j|k\rangle =0$\,.

\end{appendix}



\begin{thebibliography}{}
\bibitem{kinetic} In order not to make any confusion we here distinguish
such a general kinetic equation from the kinetic equation of
Boltzmann and the Boltzmann-like equation in both nonrelativistic
and relativistic physics. The former, derived from the many-body
Liouville equation, is exact, closed and highly nonlinear. It is
satisfied by the reduced one-body distribution function. The kinetic
equation of Boltzmann is a special case of this general kinetic
equation under appropriate approximations
\cite{Prigogine63,Balescu}. Notice that in many condensed matter and
high-energy literatures (for examples,
Refs.~\cite{Kadanoff,Litim02}) the ``Boltzmann'' equation is in fact
``Boltzmann-like'' which, typically, is the kinetic (transport)
equation of Vlasov, Fokker-Planck, Landau, or of Boltzmann but with
the scattering cross section calculated at the level of the weak
coupling (Born) approximation.
\end{thebibliography}
\end{document}